# Advances in Microfluidics and Lab-on-a-Chip Technologies

Harikrishnan Jayamohan, Valentin Romanov, Huizhong Li, Jiyoung Son, Raheel Samuel, John Nelson, Bruce K. Gale

## Abstract

Advances in molecular biology are enabling rapid and efficient analyses for effective intervention in domains like biology research, infectious disease management, food safety and bio-defense. The emergence of microfluidics and nanotechnologies has enabled both new capabilities and instrument sizes practical for point-of-care (POC). They have also introduced new functionality, enhanced the sensitivity, and reduced the time and cost involved in conventional molecular diagnostic techniques. This chapter reviews the application of microfluidics for molecular diagnostics methods like nucleic acid amplification, next generation sequencing, high resolution melting analysis, cytogenetics, protein detection and analysis, and cell sorting. We also review microfluidic sample preparation platforms applied to molecular diagnostics and targeted to sample-in, answer-out capabilities.

## Overview

Microfluidics have come a long way since the seminal paper by Andreas Manz in 1990 (Manz, 1990). The paper envisioned an integrated, automated platform for performing a range of analysis steps. The advances since then have brought us closer to the vision of a sample-in-answer-out platform (Jenkins & Mansfield, 2013; Lee, 2013). At the turn of the century, the introduction of PDMS and soft lithography gave a major boost to the field (Duffy, et al., 1998; Unger, et al., 2000). These simple, inexpensive, and rapid microfabrication techniques have enabled researches to apply microfluidics to a wide range of areas like catalysis, molecular point-of-care (POC) diagnostics remains the primary application area of microfluidics (Jayamohan, et al., 2013) and it accounts for the largest share of the microfluidics market (Yetisen & Volpatti, 2014).

Microfluidics can be broadly defined as systems leveraging micrometer scale channels, to manipulate and process low volume ($10^{-9}$ to $10^{-18}$ l) fluid samples (Whitesides, 2006). Such systems enable advantages such as the capability to process low volumes of samples requiring lower amounts of expensive reagents. Lab-on-achip (LOC) platforms leveraging microfluidics are capable of carrying out separations and detections with high resolution and sensitivity. The smaller length scales associated with microfluidics enable faster analysis and reduced response times. Advances in microfluidic manufacturing methods (lithography, xurography, laser machining) (Jayamohan, et al., 2013) have enabled devices with a smaller footprint, at a reduced cost. This is especially important for the POC applicability of microfluidic devices in the context of global health. The convergence of microfluidics with nanotechnology-based barcode techniques (quantum dots (Klostranec, et al., 2007), oligonucleotide labels (Jayamohan, et al., 2015), metal nanoparticles) has enabled multiplexed ultrasensitive detection of analytes from complex sample matrices involving contaminants (Hauck, et al., 2010; Derveaux, et al., 2008; Sanvicens, et al., 2009).

Many microfluidic platforms are limited in their application and adoption by requirements involving the need for significant off-chip sample preparation. Recent developments in on-chip sample preparation have offset some of these challenges. Microfluidic systems also suffer from challenges due to scaling like capillary forces, surface roughness, air bubbles (Lochovsky, et al., 2012), surface fouling (Schoenitz, et al., 2015), channel clogging, and laminar flow-limiting reagent mixing to diffusion. Other issues relate to volume mismatch between real-world samples and microfluidic components, and interfacing of electronics and fluids at the microscale (Fredrickson & Fan, 2004).

In spite of the significant academic interest in microfluidics, the commercial applications have not evolved at a similar rate (Chin, et al., 2012). The success of materials like polydimethylsiloxane (PDMS) in microfluidics academic research has not translated over to industry well due to issues with manufacturability and scaling. Also, there is a lack of academic research on microfluidic devices fabricated using alternative materials (glass, thermoplastic polymer), which has prevented the rapid transfer of these technologies from the lab to the market (Yetisen & Volpatti, 2014). Microfluidic commercialization is also limited due to the custom nature of each assay or microfluidic chip: there is no universal fabrication approach that can be implemented in a majority of needed applications. Another area of concern is the lack of statistical reproducibility and microfluidic chip-to-chip variability among published research (Becker, 2010).

New processes like droplet (emulsion) and paper microfluidics seems to be overcoming some of these challenges with increasing adoption by both industry and researchers alike (Lee, 2013; Hindson, et al., 2011). Droplet microfluidics utilize two immiscible fluids to establish compartmentalization within pico- or nanoliter sized droplets (Teh, et al., 2008). Paper microfluidics replaces hollow, free-flow microchannels with woven microfibers of paper that wick fluids, circumventing the need for additional pumps (Lee, 2013), but giving up some flexibility. As might be imagined, microfluidics printed on paper can be relatively inexpensive. Looking forward, 3D printing holds promise in extending these capabilities to other materials, including plastics, for microfluidic device development.

Overall, microfluidic approaches to a wide variety of molecular diagnostics applications are developing rapidly. In this chapter, we will briefly review some of the most important and most impactful applications of microfluidics in molecular diagnostics. Applications in nucleic acids, proteins, cell preparation for molecular diagnostics, and other targets will be discussed briefly.

## Microfluidics for DNA Amplification and Analysis

DNA analysis and amplification is becoming standard practice in many diagnostic and analytical procedures, with Polymerase Chain Reaction (PCR) being one of the most robust and popular molecular diagnostic techniques in medicine (Chang, et al., 2013). DNA amplification techniques can be broadly categorized as isothermal and nonisothermal. Isothermal DNA amplification techniques are carried out at constant temperature and tend to be simpler mechanically, so interest in this area is high, leading to a number of isothermal DNA amplification techniques being reported in the last couple of years (Chang, et al., 2013) (See also Chapter 3).

Isothermal DNA amplification techniques are well suited for microfluidic integration in applications where reasonably fast (15–60 minutes) DNA amplification is needed in low-resource settings, as temperature cycling is not needed, which significantly simplifies the hardware needed to carry out isothermal DNA amplification. Table 1 summarizes some of the promising isothermal DNA amplification techniques that have been successfully demonstrated in microfluidic systems in commercial and academic settings. For further details the readers can refer to reviews by Asiello and co-workers, and others (Asiello & Baeumner, 2011; Craw & Balachandran, 2012; Tröger, et al., 2015).

| Reaction | Type of template required | Reaction Temp (°C) | Highlights/Comments | Multiplex capability |
|---|---|---|---|---|
| Nucleic acid sequence-based amplification (NASBA) | RNA | 41 | 1. Prone to non-specific amplification<br>2. Requires initial heating of template RNA at 65 °C | Yes |
| Loop-mediated isothermal amplification (LAMP) | ss-DNA | 60-65 | 1. Using two primer sets, the LAMP reaction becomes very specific.<br>2. Requires careful design of primer sets.<br>3. Ease of detection of amplified products due production of pyrophosphate (visible to the naked eye) as a by-product of a positive LAMP reaction. | Yes |
| Helicasedependent amplification (HDA) | ds-DNA | 45-65 | 1. Utilizes a single primer set; which makes HDA a simple process, with ease of optimization.<br>2. However, the speed of HDA is very low when samples contain <100 DNA copies. But optimizing the reaction for a specific amplicon can compensate for this. | 2-plex |
| Stranddisplacement amplification (SDA) | ss-DNA | 37-70 | 1. Requires initial heating of template DNA at 95 °C<br>2. Prone to non-specific amplification 3. Slow reaction | Yes |
| Recombinase polymerase amplification (RPA) | ds-DNA | 37-42 | 1. Fast reaction (probably one of the fastest among other isothermal DNA amplification techniques)<br>2. A robust reaction; without | Yes |

| | | | requiring precise temperature control | |
|---|---|---|---|---|

Table 1: Promising isothermal DNA amplification techniques for incorporation in microfluidic systems.

PCR is the predominant and most popular non-isothermal DNA amplification technique and used in many microfluidic devices. PCR involves three sub-steps that occur at different temperatures. PCR typically requires at least 35 to 40 temperature cycles for a single PCR reaction to achieve useful concentrations. The speed at which PCR can be run is dependent on two factors: the speed of the DNA polymerase and the heat transfer rate of the hardware performing the PCR. As microfluidic systems are inherently small, leading to a small thermal mass, and having a high surface-to volume ratio, they are naturally capable of providing rapid heat transfer rates. Microfluidic PCR systems also offer the ability to automate the preparation of the PCR reaction mix, thereby reducing the risk of contamination and false positives by human error. Finally microfluidic PCR systems require low sample volumes, which are helpful when the genetic material being tested is scarce, and the reagent volumes are likewise low, significantly reducing costs.

Since the inception of microfluidics in the 1980s a considerable amount of work has been done to develop microfluidic devices for PCR. There are generally two types of microfluidic systems for PCR: flow-through PCR, stationary PCR, and droplet digital PCR (Chang, et al., 2013). In a flow-through PCR system, the PCR mixture travels through a microchannel that contains temperature regions for all three sub-steps of PCR. In some versions, the sample may be moved back and forth between the temperature regions while in others the sample reaches the temperatures by continually moving forward. In stationary PCR systems the PCR mixture remains stationary in a microchamber while the temperature of the microchamber cycles through the needed temperatures. There are many variations of these approaches. For example, in droplet digital PCR systems, the PCR reaction mix along with the template DNA is encapsulated in a microdroplet and then transported to different regions of a microchip or temperature cycled in place (Prakash, et al., 2014).

Thousands of microfluidic PCR devices have been successfully demonstrated with measurable real-time amplification incorporated in the microfluidic PCR chip and some show amplification completed in a few minutes (Chang, et al., 2013; Pješčić, et al., 2010; Crews, et al., 2008; Neuzil, et al., 2006), even at the single cell level (Zhu, et al., 2012). For example, Figure 1 shows a microfluidic chip performing both continuous PCR and high resolution melting analysis (HRMA) simultaneously in less than 6 minutes for 30 cycles. The PCR is progressing down the image while HRMA can be performed simultaneously for each cycle in the horizontal direction by measuring the fluorescence intensity in the image. A similar chip has recently been show to complete PCR in less than 1 minute (Samuel, et al., 2016) using extreme PCR (Farrar & Wittwer, 2015). Furthermore in the last 5 years, biomedical diagnostic companies have commercialized several microfluidic PCR systems (Cao, et al., 2015; Volpatti & Yetisen, 2014).

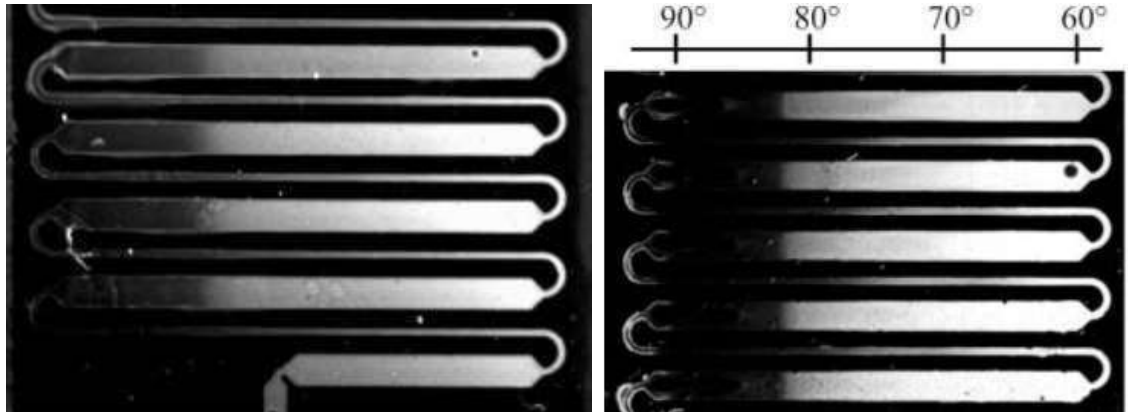

Figure 1. Spatial Continuous Flow PCR showing both PCR and HRMA melting [27]. (Left) The image shows the microfluidic chip design that has two types of channel widths. Denaturation of DNA occurs in narrow channels, while annealing and extension (which are relatively slower than denaturation) occur in the wider channels due to reduced flow rate. (Right) The temperature zones along the chip are labeled (°C).

## DNA Sequencing and Mutation Detection

In cancer and other diseases, altered DNA gene patterns or mutations have been found to be useful biomarkers for detection and diagnosis of disease (Almoguera, et al., 1988). Detecting mutations requires the ability to sequence at least a small part of a genetic sequence, which has led to major efforts to develop high-speed, highthroughput DNA sequencing methods. As microfluidics has emerged as a tool for clinical molecular diagnostics, applications in mutation detection and genetic screening have developed with the promise to profile genetic sequences quickly and to interpret the implication of such sequences. Traditional macro DNA sequencing includes steps such as cell preparation, amplification, purification, and electrophoresis. Each step can be scaled down and integrated into a microfluidic device to achieve rapid and low-cost DNA sequencing (Paegel, et al., 2003). Other nontraditional approaches, often adapting macroscale methods, for detecting altered gene sequences or sequencing short sections of genes have been developed, including: digital PCR, and HRMA. The application of microfluidic technology to many of these sequencing or mutation detection techniques is discussed below.

## Capillary Electrophoresis

In 1995, Wooley and co-workers developed a microfabricated capillary electrophoresis (CE) chip, which can complete DNA sequencing with 97% accuracy and ~150 bases in 540 s for four-color separations. The CE chip demonstrated the feasibility of fast and high-throughput DNA sequencing (Woolley & Mathies, 1995). In 1999, Liu and co-workers presented an improved microfabricated CE chip. The separation matrix, temperature, channel and injector size, and injector parameters were all optimized to achieve better DNA sequencing performance. The optimized chip could achieve ~500 bases in 20 min for four-color separations (Liu, et al., 1999). Paegel and co-workers developed a radially symmetric, 96-lane capillary array electrophoresis chip, which acquired ~41000 bases in only 24 min (Paegel, et al., 2002). Similar approaches and further improvement studies were summarized in (Paegel, et al., 2003) and these approaches are regularly used in recent efforts. This topic was reviewed in depth in a previous version of this book (Jayamohan, et al., 2013). As an example of what has been accomplished recently, micro CE integrated systems have been used for quantitative detection of low-abundance mutations of the KRAS gene from paraffin tissue sections of colorectal cancer. These systems have nano-liter sample introduction components leading to CE separation of the target genes and detection by laser-induced fluorescence (LIF); all being accomplished in minutes or even seconds (Zhang, et al., 2013; Xu, et al., 2010).

## DNA purification

He and co-workers used capillary zone electrophoresis for purification of sequencing fragments (He, et al., 2000). Khandurina and co-workers developed a microfluidic device for fraction collection of various size DNA fragments (Khandurina, et al., 2002). In Tian's study, the effectiveness of a variety of silica resin for miniaturized DNA purification was evaluated (Tian, et al., 2000). Other alternative on-chip approaches for DNA purification were also studied, including using temperature gradients along the channel and hybridization-mediated capture (Paegel, et al., 2003). More recent techniques not specifically for sequencing, but generally relevant are discussed in the sample preparation section.

## DNA amplification and Sanger sequencing

Many studies performing low volume Sanger cycle sequencing in microfluidics have been presented. Hadd and co-workers, and Xue and co-workers both demonstrated low volume reactions inside a capillary, which established the feasibility of smallscale sample preparation (Hadd, et al., 2000; Xue, et al., 2001). Lagally and coworkers developed the first nanoliter-scale DNA amplification systems, which was integrated with electrophoretic analysis on a microfluidic chip (Lagally, et al., 2000). In 2006, Blazej and co-workers developed a microfabricated bioprocessor to integrate all three Sanger sequencing steps. This micro device was built on a hybrid glass-PDMS wafer and enables complete Sanger sequencing from 1 fM of DNA template within 30 min. With further improvements, the starting template for DNA sequencing has been reduced from 1 fM to 100 aM (Blazej, et al., 2006). In 2013, Abate and co-workers developed a droplet-based microfluidic system for DNA sequencing in a rapid and inexpensive manner (Abate, et al., 2013).

## Digital PCR

While PCR has long been used for mutation detection, a highly sensitive version of PCR, digital PCR, has been gaining a significant following both commercially and scientifically. Droplet-based digital PCR puts the PCR milieu into thousands or millions of drops with a general goal of keeping any amplification targets at less than 1 per drop, which results in an "on" of "off" signal for each drop when the PCR is complete. Digital PCR can reduce overall analysis costs and reduces sample cross contamination. Most importantly droplet-based detection can provide a highly sensitive and high-throughput method for detecting DNA mutations (Hsieh, et al., 2009; Pekin, et al., 2011). As an example, droplet-based digital PCR enabled the precise determination of mutations in several cancer cell-lines and the precise quantification of a single mutated KRAS gene in a background of 200,000 unmutated KRAS genes (Pekin, et al., 2011). In a similar vein, BRAF mutation detection was accomplished using a spinning disk digital PCR layout (Figure 2) (Sundberg, et al., 2014). In this case a polymer microfluidic system the size of a DVD was fabricated with a single spiral channel that contained 1000 microwells on the outside of the disc. After injecting a PCR reaction mix, the disc is spun and a combination of centrifugal forces and the laminar flow profile distribute the reaction mix into the wells. An oil plug is then flowed through the disk to isolate the wells. At that point conventional thermal cycling can be performed to achieve digital PCR and real-time detection of product based on fluorescence in each of the wells. The main advantage/distinction of this system is that by using a simple setup a relatively large reaction mix is distributed into nanoliter sized volumes automatically.

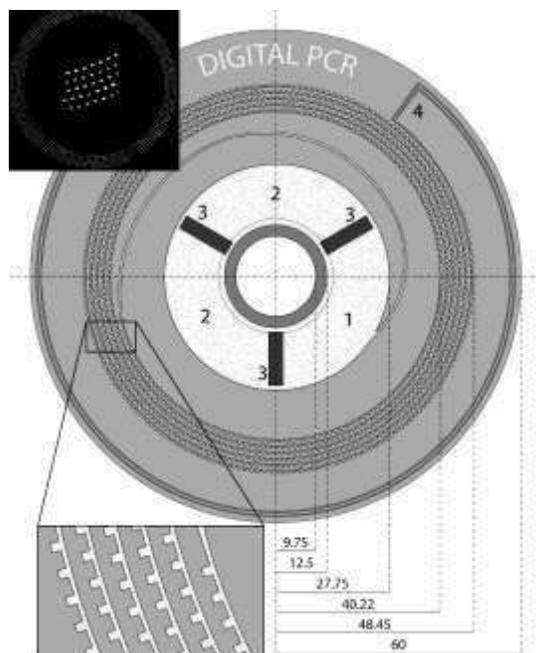

Figure 2. Spinning disk for digital PCR. (Inset) Post PCR fluorescent image of the disk. Fluorescence correlates to number of copies per well. Reprinted with permission from (Sundberg, et al., 2014). Copyright 2010 American Chemical Society.

## Microfluidics for High Resolution Melting Analysis (HRMA)

HRMA is performed after DNA amplification is completed and is focused on the amplified region of the DNA strand (See also Chapter …). In HRMA, intercalating dyes are incorporated into the double-stranded DNA (ds-DNA). These intercalating dyes fluoresce brightly when incorporated in ds-DNA, but fluoresce poorly when that dsDNA becomes single-stranded DNA (ss-DNA), which occurs when a ds-DNA melts into ss-DNA as the temperature is increased past the melting temperature (usually from 50ºC to 95ºC). The melting point of ds-DNA is very sensitive to DNA sequence and any mismatches, and there is a measurable drop in fluorescence as the amplicon melts. After the melting procedure, a "melt curve" is obtained by plotting fluorescence intensity as a function of temperature. The melt curve profile is unique for a particular DNA sequence (even down to a single base in the DNA sequence); hence by analyzing the melt curve profile one can identify variations in a DNA sequence.

Implementation of HRMA requires a way to change the temperature of the sample and a way to measure the fluorescence output, both of which are readily achieved at the microscale, and the thermal benefits of microfluidics apply to HRMA as well. Microfluidic systems for HRMA can be distinguished based on how the temperature gradient needed to obtain the 'melt curve' is developed. There are currently two methods to develop temperature gradient in microfluidics: temporal melting or spatial melting (Crews, et al., 2009).

Temporal melting is the conventional way to develop a temperature gradient in macroscale systems and has been replicated in microfluidic systems. Temporal melting is done in a fixed reservoir containing the PCR product and is basically a gradual heating process. In this case the slow heating rates are crucial for accuracy and sensitivity of the melt curve, but the slow rate of heating makes temporal melting not suitable when fast and robust HRMA is desired. A number of microfluidic devices have been made in the last decade that utilize some form of temporal DNA melting for DNA identification. Most configurations involve a heating element incorporated internally into the microfluidic chip or externally outside the chip's body. The heating is done either by thermoelectricity or an external heat source like lasers to generate a temporal thermal gradient (Lee & Fan, 2012; Athamanolap, et al., 2014). Multiple images of the reservoir are taken to monitor the change in fluorescence with change in temperature to generate a melt curve.

Spatial melting is achieved by establishing a temperature gradient across an elongated reservoir and is only possible in microfluidic systems. When the elongated reservoir is filled with a PCR product, a spatial variance of fluorescence is observed along the elongated reservoir. A single image of the reservoir is taken and the melt curve is generated from it. In spatial melting the melting can be performed on either flowing or stationary fluids, as it is not dependent on time and is best suited for fast HRMA. We have reviewed (see Table 2) significant work reported in the literature on microfluidic HRMA utilizing spatial melting.

| **Publication title** | **Highlights/Comments** |
| --- | --- |
| Product differentiation during continuous-flow thermal gradient PCR (Crews, et al., 2008) | Melting analysis is the main focus of this publication. The authors show how the performance of their device to carry out fast PCR/HRMA compares with commercial equipment. |
| Glass-composite prototyping for flow PCR with in situ DNA analysis (Pješčić, et al., 2010) | Melting analysis is not the main focus of this publication; however the authors demonstrate how PCR and HRMA can be combined and carried out on a single microfluidic chip |
| Spatial DNA melting analysis for genotyping and variant scanning (Crews, et al., 2009) | 1. Spatial microfluidic HRMA is used for SNP scanning and genotyping.<br>2. HRMA is shown in a continuous-flow regime.<br>3. Up to 20 samples were processed for HRMA in serial fashion in the same device without any cleaning steps in-between. |
| Real-time damage monitoring of irradiated DNA (Pješčić, et al., 2011) | The authors demonstrate real-time measurement of DNA damage due to radiation exposure using a microfluidic HRMA |
| Genotyping from saliva with a one-step microdevice (Pješčić & Crews, 2012) | PCR and spatial HRMA are carried out on a single disposable microfluidic chip and shown to distinguish between human male and female saliva samples. |

Table 2. Publications reporting significant progress in spatial melting of DNA for microfluidic-enabled HRMA.

### DNA Methylation Detection

DNA methylation, the covalent addition of a methyl group to the cytosine base in DNA, is a central epigenetic modification and has an essential role in cellular process including genome regulation, development and disease. Microfluidics has been shown to improve the DNA methylation detection process, improve efficiencies in time and cost by making the analysis high-throughput and sensitive with small sample volumes (Paliwal, et al., 2010).

One of most well-known microfluidic DNA methylation detection methods is using a platform with an array of microfluidic channels and an array of chambers (Weisenberger, et al., 2008). The approach was tested with single methylated *PITX2* molecules. After the sample was amplified in multiple PCR reaction wells, individually amplified methylated DNA molecules were then visualized via probe fluorescence signals using a high-resolution CCD camera. This method was able to successfully and sensitively detect single molecule DNA methylation events in a small PCR reaction volume.

Another notable approach is using methylation-specific PCR (MSP) within an arrayed micro droplet-in-oil platform. The device has nine snowflake-like functional units arranged in a circular array (Figure 3). Each function unit consists of 12 open reaction chambers, which are also arranged in a circular array and connected to a sample access port through a microfluidic network. Methylation specific primers are predeposited into reaction chambers. The device can perform 108 MSP reactions simultaneously. Each functional unit is capable of DNA methylation analysis of multiple genes with single sample dispensing, thereby significantly reducing the sample preparation time, improving throughput and allowing for automation. This method uses mineral oil as a working fluid for actuation, preventing contamination and evaporation of the sample. This method is exemplified by analysis of two tumor suppressor promoters, *p15* and *TMS1* (Zhang, et al., 2009).

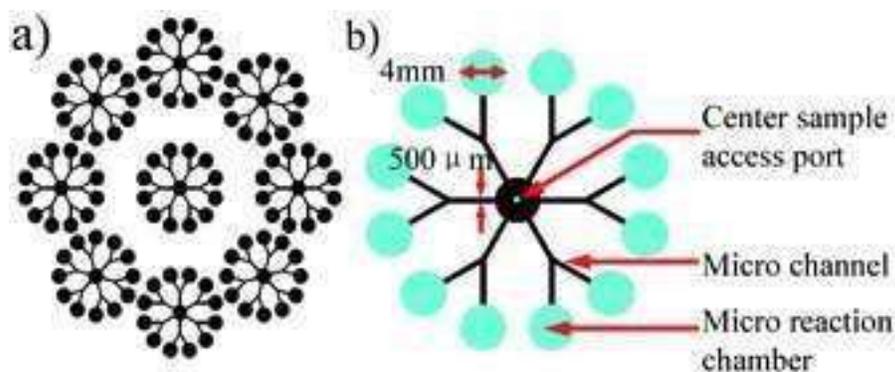

Figure 3. a) Overall layout showing nine snowflake-like functional units arranged in a circular array. (b) Individual functional unit layout. Reprinted with permission from (Zhang, et al., 2009). Copyright 2009 American Chemical Society.

## Padlock and Selector Probes

Both padlock probes and selector probes are linear oligonucleotides with two complementary ends to the target DNA strand for DNA analysis and molecular diagnosis (Jansson, 2007) (See also Chapter ...). Current molecular diagnostic approaches need manual analysis by skilled personnel, which is time-consuming and labor-intensive. The application of microfluidic technology has been increasing because of the small sample volume usage, easy portability, and rapid detection time (Tröger, et al., 2015). A few examples of microfluidic devices utilizing padlock and selector probe technologies are described below.

In 2005, Melin and co-workers developed a thermoplastic microfluidic platform for multiple purposes: sensitive detection, cell culture and actuation. Padlock probes/rolling circle amplification (RCA) was employed in a platform for DNA detection (Melin, et al., 2005). In 2006, Jarvius and co-workers developed an approach for quantitative single-molecular detection based on padlock probe ligation using a microfluidic system. This method was applied to sensitive detection of the bacterial pathogen *Vibrio cholerae* (Jarvius, et al., 2006).

In 2008, Mahmoudian and co-workers developed an integrated platform, on which both Circle to Circle Amplification (C2CA) and RCA were successfully performed with padlock probes. The microchip is made from poly(methyl methacrylate) (PMMA). 25 ng of bacterial genomic DNA was detected within 65 min (Mahmoudian, et al., 2008). In 2010, Sato and co-workers created an integrated microchemical chip and combined the padlock probe and RCA on chip. The microchip was made from glass and contains Y-shaped channels with a dam structure. 88 ng Salmonella genomic DNA was detected using on-bead RCA on a microchip (Sato, 2010).

In 2011, Konry and co-workers demonstrated a droplet-based PDMS microfluidic device to detect protein markers based on padlock probe technology and RCA methods. After highly specific antigen-antibody recognition, less than 10 EpCAM surface tumor markers per cell was successfully detected with visual fluorescence (Konry, et al., 2011). Ahlford and co-workers presented a microfluidic system for DNA analysis of KRAS using a highly selective padlock probes and C2CA (Ahlford, et al., 2011). Tanaka and co-workers used a glass microchip for DNA detection based on RCA methods with padlock probes. DNA detection in small volume samples was achieved (Tanaka, et al., 2011).

In 2014, Kühnemund and co-workers demonstrated a digital microfluidic (DMF) chip to perform C2CA with a padlock probe. The microchip is made from glass and integrates all the assay steps except for heating. A novel magnetic particle shuttling protocol was employed to enable high-sensitivity DNA detection (Kühnemund, et al., 2014). Mezger and co-workers developed a rapid and sensitive microfluidic PDMS chip to detect highly variable dsRNA viruses using padlock probes (Mezger, et al., 2014). Sato and co-workers developed an automated microfluidic system using RCA methods to simplify the single DNA counting process in a cell (Kuroda, et al., 2014).

It is worth noting that in some cases, where microfluidic automation is not ideal, the integration of samples can be a complicated task. Besides the potential manual steps involved, the application of padlock and selector probes technologies on chip has largely improved the DNA detection and analysis efficiency with small volume samples. Due to the high sensitivity and the reduced sample requirements, this technology can be a powerful tool in diverse diagnostic fields.

## Microfluidics in Cytogenetics

In the field of cytogenetics, there are various techniques such as Fluorescence InSitu Hybridization (FISH) assays that require expensive reagents, a large time commitment, and a need for experienced and well trained technicians (Kwasny, et al., 2012)(See also Chapter …). This makes such techniques less favorable in many cases even if they may provide better results. LOC devices have been designed to improve the efficiency of many cytogenetic techniques, allowing for much faster and more reliable results (Kwasny, et al., 2012).

### FISH analysis

Traditional FISH analysis involves a long and complex protocol, which leads to detection of genetic abnormalities. The probes used to visualize the presence of a DNA sequence have high costs, requiring more than $100 for every individual test (Kwasny, et al., 2012). LOC devices have been designed which make this process much more efficient by decreasing the time commitment, decreasing the amount of probes and sample needed, and by automating the process to a greater degree, thereby achieving consistent results. Perez-Toralla and co-workers showed that their device was capable of decreasing the volume of sample and probes needed for FISH by a

factor of 10, while simultaneously cutting the time required in half. This device could be fully automated and obtained the same quality results as a traditional protocol (Perez-Toralla, et al., 2015). Other devices have been able to demonstrate similar improvements, allowing for up to 96 samples to be analyzed simultaneously using the same volume of probe that would usually be used for 1 test (Kwasny, et al., 2012). In general, LOC devices used in FISH analysis can reduce the time invested, cost of reagents used, and automate processes that would otherwise require extensive training and experience. However, some of the most efficient devices are rather complicated and expensive to manufacture, which may limit the benefits of reducing the cost of reagents.

## Microfluidics for Protein Detection and Analysis

Human blood plasma contains an enormous amount of proteins, numbering around $10^{10}$ (Jacobs, et al., 2005). This coupled with recent research demonstrating that blood plasma also plays host to critical disease biomarkers such as exosomes (Kalra, et al., 2013) and miRNA (Mitchell, et al., 2008), paints a picture of a highly complex sample. Detection systems that can effectively and rapidly identify and analyze proteins from such mixtures can greatly enhance molecular diagnostic capabilities with downstream benefits in personalized health care (Gonzalez de Castro, et al., 2013).

A variety of microfluidic systems have been developed from existing macroscale techniques for the identification and analysis of proteins. Typically, traditional protein identification approaches are derived from one of two popular methods, immunoassays or immunoblotting. Immunoassays are based on the interaction of antibodies, whether adsorbed to a surface or in free solution with specific antigens (Ng, et al., 2010). Immunobloting on the other hand is used to first determine the molecular mass of the protein via electrophoresis gel migration before incubation with antibodies for detection and identification (Hughes & Herr, 2012). Both methods suffer from diffusional limitations, excessive reagent consumption, reproducibility, and throughput restrictions (Jin & Kennedy, 2015). Microfluidics can help reduce the diffusional distances by increasing surface area to volume ratios, reduce reagent consumption through micro- and nano- fabricated channels and chambers, and automate all steps of the process (Ng, et al., 2010).

Traditional methods of protein analysis typically involve the implementation of two strategies for the analysis and sequencing of proteins based on mass spectrometry (MS): matrix assisted laser deposition and ionization (MALDI) or electrospray ionization (ESI) (Domon & Aebersold, 2006). ESI utilizes a small nozzle or a capillary to drive reagents into the mass spectrometer for analysis (Figure 4a) by reducing the charged droplets into molecular ions (Han & Gross, 2005). Alternatively, MALDI utilizes a dry crystalline matrix to affix a sample for laser interrogation (Figure 5a). The crystalline matrix helps to desorb and ionize the sample resulting in the sublimation of the protein species leading to the formation of charged ions (Aebersold & Mann, 2003; Hardouin, 2007), which can then be analyzed by MS. One of the main challenges of using ESI-MS is the suppression of ions due to high salt concentrations in the buffer (Gao, et al., 2013). This makes analysis very difficult or near impossible. Further, while MALDI-MS is perfectly capable of analyzing high salt buffers, the structural matrix makes analyzing low mass structures difficult due to noise generation in the resulting spectra of the sample (Gao, et al., 2013).

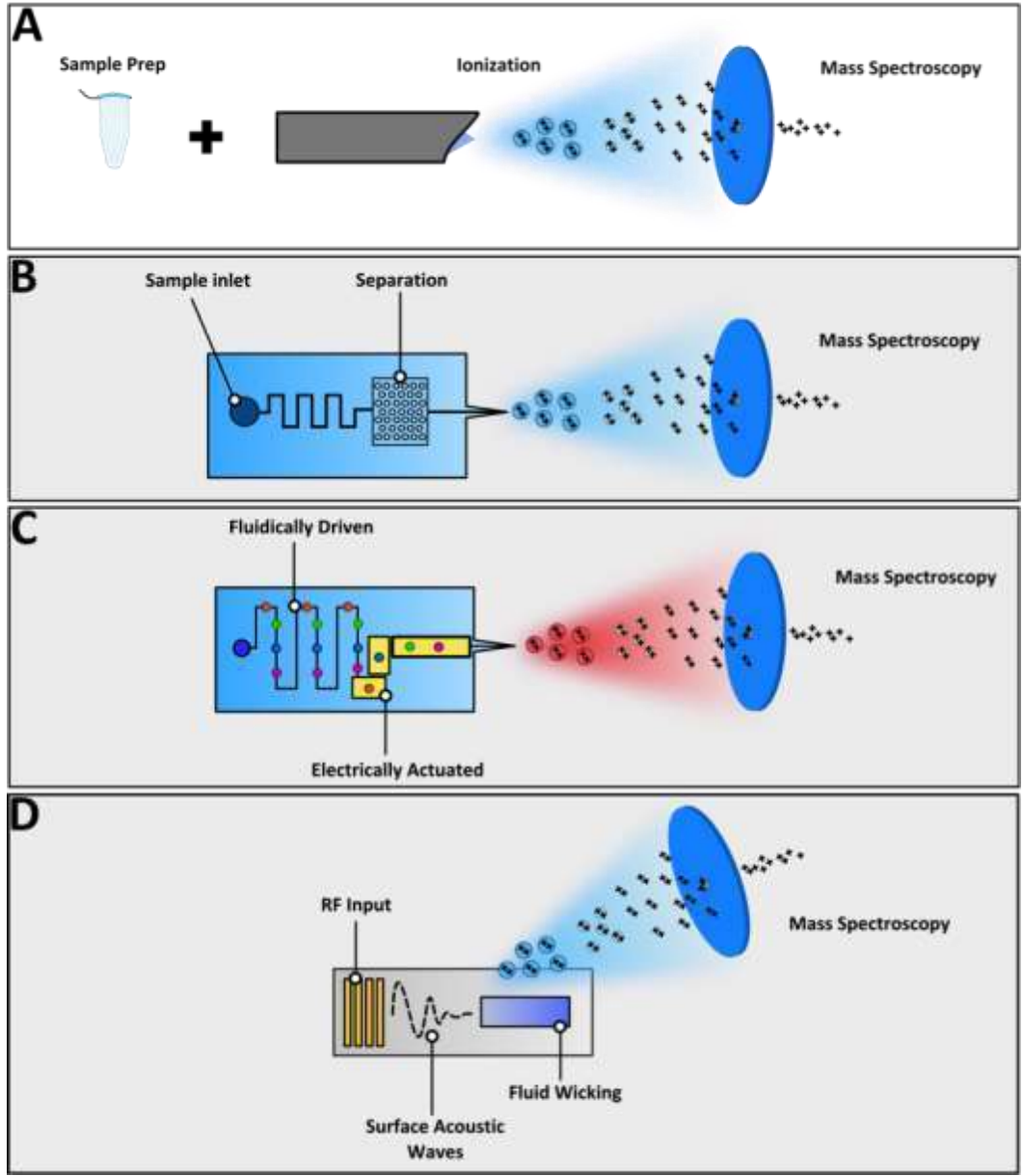

Figure 4. a) Traditional approach to ESI-MS. Sample is processed before ionization and MS analysis, b) Microfluidic network where on chip separation and processing leads directly to ionization and MS analysis (Mery, et al., 2008; Sainiemi, et al., 2012), c) DMF can utilize either pressure driven or electrically actuated droplet transfer to an ionization site for MS analysis (Shih, et al., 2012; Baker & Roper, 2012; Ji, et al., 2012), d) Paper based microfluidic approach coupled with surface acoustic waves for ionization and MS analysis (Ho, et al., 2011). Each figure was designed to give a general idea of the methods discussed and techniques employed and are not necessarily representative of the actual devices.

The associated challenges with using MS and the aforementioned preparatory techniques can be readily addressed with microfluidics. Miniaturization leads to smaller sample volumes, reduced diffusional distances, the ability to carry out high throughput analysis, system automation, parallelization and process streamlining of processes (Chao & Hansmeier, 2013).

## Overview of ESI-MS integrated microfluidic platforms

Several methods have been proposed for chip fabrication for downstream integration with MS including traditional wet etching techniques (Lazar, et al., 2006), surface micromachining (Xie, et al., 2005) and rapid prototyping (Yin, et al., 2005). The major features of the device are etched from a silicon substrate via photoresist deposition, followed by PDMS curing and bonding to either a silicon substrate or a glass slide. Electrical contacts can be added at anytime during the fabrication process via evaporation or sputtering. Further complexity can be added to the microfluidic device by integrating valves, gates and chambers for eliminating fitting, leaking and blocking issues (Srbek, 2007).

Traditionally, on-site filtration occurs through an area packed with microbeads. Loading the beads can be quite a challenge at the microscale. Vinet and co-workers demonstrated a robust method for fabricating 2-D ordered arrays of silicon micropillars (Figure 4b) using deep reactive ion etching (DRIE) on a silicon substrate for effective sample separation, as an alternative to microbeads and precise nozzle fabrication (Mery, et al., 2008). Using a tryptic peptide mix they were able to show effective separation and stable electrospray operation. In a separate study, Ketola and co-workers modified the surface of their structures with C-18 for reverse-phase separation or $SiO_2$ for normal separation. Further, by integrating a silicon base with a glass cover they were able to fabricate a 3D ESI tip while allowing for microfluidic chip operation with both laser-induced fluorescence and MS (Sainiemi, et al., 2012). The system required only 10nL of reagents, demonstrating fast separation and good sensitivity. One key limitation of this approach is the need for microfabrication of the pillar arrays. Rigidity and resolution of the features are material and process dependent, while reusability of the devices due to clogging may also be an issue.

In an entirely different approach, Shih and co-workers successfully demonstrated the integration of DMF with ESI-MS analysis. An electrical potential was used to drive droplets closer to microcapillary acting as a directly integrated ESI tip (Figure 4c). Sample uptake occurred through capillary action where an applied DC voltage was used to generate the spray interface (Shih, et al., 2012). The device was successfully used for the identification of a specific marker in a dried blood spot sample. Similarly, Baker and Roper utilized a capillary and an eductor to transfer droplets from an open or closed setup to the ESI tip. In conjunction with nitrogen, an eductor was used for generating areas of negative pressure at the ESI tip via the Venturi effect triggering droplet movement (Baker & Roper, 2012). Continuous, high throughput analysis of the entire droplet volume either inside the device or in ambient air is also a possibility. Further, Ji and co-workers utilized DMF for rapid proteolysis (Figure 4c) by encapsulating and digesting fractions inside droplets (Ji, et al., 2012). The advantage of this approach is reduced cross contamination, sample loss and nonspecific absorption. One of the key limitations is in the design and fabrication of the electrical circuits to drive the process. Further, as with any ESI tip, clogging can be an issue.

As mentioned earlier, a potential disadvantage of traditional approaches to microfluidic device fabrication is the need for clean room access, materials and expertise. Paper-based microfluidic devices are low cost, biodegradable, transportable and effective at delivering samples to the site of interest (Mao & Huang, 2012). Ho and co-workers demonstrated a paper based microfluidic device that utilizes wicking (Figure 4d) for effective sample uptake from a reservoir for MS analysis. Surface acoustic waves were used to ionize the filtered sample at the end of the paper, effectively demonstrating the ability to process both high ionic and viscous samples, conditions that may prove difficult for traditional ESI nozzles to accommodate (Ho, et al., 2011).

## Overview of MALDI-MS integrated microfluidic platforms

Microfluidics can be integrated with the MALDI framework in several ways. The first approach utilizes the manipulation of a sample in a microfluidic reactor for direct deposition onto a MALDI plate. The second method uses the microfluidic device as the MALDI plate where the sample can be prepared on site and directly inserted into the MALDI-MS instrument for analysis (DeVoe & Lee, 2006; Lee, et al., 2009).

DMF is a popular platform for integration with MALDI-MS. Chatterjee and coworkers demonstrated a microfabricated device capable of efficiently processing proteins via droplet manipulation (Figure 5b) and drying, ready for MALDI interrogation (Chatterjee, et al., 2010). Disulfide reduction, alkylation and enzymatic digestion were carried out within the device consisting of a removable top lid and a bottom plate with integrated electrodes. Further droplet control was demonstrated by electrowettingon-dielectrics (EWOD) with in-line sample purification for deposition onto a stainless steel target for MALDI analysis (Wheeler, et al., 2005). In a three stage process, the sample containing peptides and impurities was moved and dried, impurities dissolved and removed by a second droplet, and MALDI matrix deposited on top via a third droplet (Figure 5b).

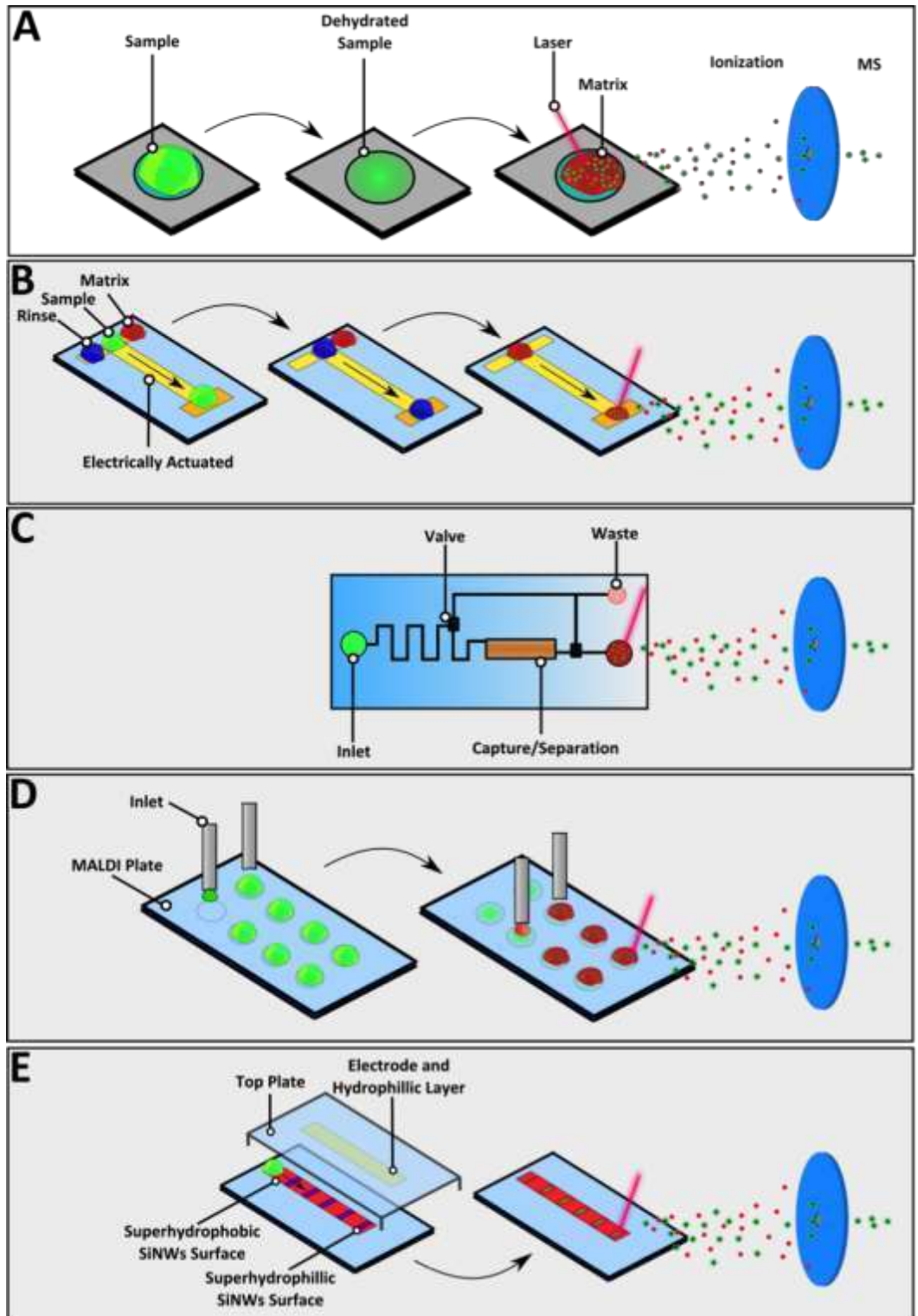

Figure 5. a) Traditional approach to MALDI-MS. A sample is added to the plate, followed by drying, matrix deposition and laser interrogation, b) DMF can be used for moving droplets that contain the sample, the rinsing and the matrix. The matrix can either be deposited manually by removing the top lid (Chatterjee, et al., 2010) or by moving a droplet in place (Wheeler, et al., 2005), c) Pressure (Lazar & Kabulski, 2013) or centrifugally (Thuy, et al., 2010) driven flow can be used in conjunction with chromatography columns for separation, preparation and matrix deposition, d) Microfluidic devices can be used for contact and non-contact deposition of the target onto a MALDI plate. The matrix can be loaded either before or after sample deposition (Küster, et al., 2013; Ro, et al., 2006), E) Utilizing hydrophilic capture regions, the matrix step can be removed allowing for direct laser interrogation on the chip (Lapierre, et al., 2011). Each figure was designed to give a general idea of the methods discussed and techniques employed and are not necessarily representative of the actual devices.

Further, on-chip MALDI processing was effectively demonstrated by Lazar and Kabulski. Electroosmotic pumps were integrated with a liquid chromatography channel for sample separation before analysis (Lazar & Kabulski, 2013). Valves were used to control the flow of sample through the slurry-loaded separation channel (Figure 5c). The sample was prepared by manual addition of the MALDI matrix for analysis, successfully displaying fM sensitivity for bovine cytochrome C and hemoglobin. Thuy and co-workers followed a different approach wherein a CD based microfluidic device was used to prepare, digest and analyze a sample all in a single, automated run (Thuy, et al., 2010). Centrifugal force was used to drive the sample through an affinity column (Figure 5c) where the protein was captured and then tryptically digested. The digest was then eluted and co-crystalized with a MALDI matrix in one of the 54 designated reservoirs. While droplet manipulation and in-line processing within the confines of the device offers several degrees of control over the process parameters, the main issue is still the need for a matrix than can lead to reduced signal to noise ratios.

Off-chip applications are also quite effective. Kuster and co-workers developed a Tjunction microfluidic device (Figure 5d) that generated nanoliter droplets guided into a capillary for deposition onto a MALDI matrix covered plate (Küster, et al., 2013). The advantage of this technique is the high throughput droplet generation potentially analyzing 26000 droplets in a streamlined process using a detection system for automated stage movement. Further, Ro and Knapp demonstrated a microfluidic device integrating an array of UV-polymerized methacrylate monolithic columns within the microfluidic channels (Figure 5d) for separating tryptic digested proteins from a peptide mixture (Ro, et al., 2006). The vertically mounted device deposited droplets of the sample onto a MALDI-MS plate for analysis.

An inherent problem with using a matrix to crystallize the sample for analysis is nonhomogeneity in the formation of the crystal leading to hot spots and reduced resolution (Northen, et al., 2007). Lapierre and co-workers demonstrated a DMF device that manipulates small amounts of volume for matrix-free laser interrogation (Lapierre, et al., 2011). A small droplet containing the sample was actuated along a channel patterned with superhydrophobic and superhydrophyllic areas (Figure 5e) on top of a silicon nano-wire interface that captured some of the liquid. Upon drying and laser interrogation, the silicon nano-wire interface acts as an inorganic target allowing for MS analysis. While highly sensitive and matrix free, the key limitation of this approach is the associated complexity in the fabrication of the device.

### Overview of popular microfluidic detection platforms

One of the strengths of microfluidics is direct customization of most traditional immunoassay protocols. Several strategies have been proposed for enhancing limits of detection, including gold nanoparticles, which act as nanoelectrodes with high electrical conductivity and surface area for antibody attachment and detection (Mani, et al., 2009). Quantum dots, due to their tunability, brightness, high absorption coefficients and photostability, have also yielded highly sensitive results (Li, et al., 2010). In a different approach, Karns and Herr utilized electrophoretic immunoassay separation of endogenous tear protein biomarkers to obtain mobility and immunoafinity information from 1 µL samples (Karns & Herr, 2011).

Fast, efficient, high throughout platforms may allow for enhanced sample quantification and, as a result, better treatment strategies. Protein microarrays have contributed a great deal towards the realization of this goal by utilizing pin printing, microstamps or micro flow printing assays (Romanov, 2014). Simple, cost effective, high throughput microfluidic devices with high sensitivity have also been described for rapid diagnoses of HIV and syphilis (Chin, et al., 2011). Selecting the correct microarray platform is crucial as platforms may vary in the types of molecules that they can print, the quality of the spots, the throughput and operational requirements.

In addition, DMF has also been utilized as miniaturized immunoassay reactors. In these systems, protein detection typically relies on optical methods; however electrochemical detection has also been demonstrated (Shamsi, et al., 2014). One of the advantages of DMF is the generation of highly tunable droplets. Vergauwe and coworkers demonstrated a highly sensitive EWOD system capable of both heterogeneous and homogeneous immunoassays by droplet manipulation (Vergauwe, et al., 2011). One of the issues with using DMF for immunoassays is sample recovery. While detection has been thoroughly demonstrated, recovering protein for subsequent study is still challenging.

Microfluidics has also been effectively utilized for western blotting. Herr's group has dedicated a significant amount of effort in improving most facets of traditional western blots including completely automated western blots with reusable chips (He & Herr, 2010) and fully integrated, rapid lectin blotting through the removal of SDS from resolved protein peaks via photopatterned microfilters within the microfluidic device (He, et al., 2011). A similar approach was also used for analysis of human sera and cell lysate utilizing a glass microfluidic chip resulting in rapid operation, on the order of 10 to 60 minutes with pM detection limits (Hughes & Herr, 2012). While cleaning and reusability of such devices was demonstrated, it's not clear how many times these devices may be regenerated before contamination or material degradation becomes an issue.

An electrostatic immobilization gel was developed as an alternative to the sandwich format typically used within microfluidic western blots. The result was a reduction of reagents consumption on the order x200 and a reduction in assay duration by 12x achieved through charge interactions (Kim, et al., 2012). As an alternative to introducing samples into a channel for separation and immobilization, Jin and coworkers fabricated a microfluidic chip for direct deposition of sieve separated protein bands on a perpendicularly mounted PVDF (Polyvinylidene fluoride) membrane (Jin, 2013). Using this approach they demonstrated reliable, reusable and reproducible separation and multiple injections using the same channel and capture membrane.

## Microfluidic Sample Preparation

There have been significant advances in application of microfluidics for molecular diagnostics applications. Extensive research has gone into the integration of molecular analysis systems (nucleic acids, proteins, pathogens, and small molecules) on-chip. These have led to a reduction in the costs of reagents and user interaction with the instruments. These platforms have leveraged advances in technologies such as: PCR, CE, FISH, surface plasmon resonance (SPR), surface enhanced raman scattering (SERS), and giant magnetoresistance (GMR), electrical/electrochemical/mechanical detection among others (Kim, et al., 2009). All these methods typically require some form of off-chip sample preparation (SPrep). Unfortunately, advances in on-chip SPrep have been overlooked in comparison to downstream processes like analysis and sensing (Brehm-Stecher, et al., 2009; Mariella Jr, 2008; Kim, et al., 2009). The dependence on traditional offchip sample pre-treatment involving expensive equipment and trained personnel has prevented the translation of these advances to POC (Byrnes, et al., 2015).

Sample preparation steps include cell lysis, washing, centrifugation, separation, filtration, and elution. These techniques performed using the conventional route are highly labor intensive, time consuming, involve multiple steps, and require expensive laboratory equipment (Byrnes, et al., 2015). For instance, nucleic acid (NA) extraction involves multiple steps to collect DNA or RNA from raw samples such as whole blood, urine, saliva etc. On-chip integration of these steps can help in reducing the total analysis time. The lower sample and regent consumption in microfluidic systems enables a lower cost of analysis. Also an enclosed sample-in, answer-out system, reduces the chance for cross-contamination. On-chip sample integration typically involves adaptation and modification of conventional macroscale laboratory methods to fit microfluidic formats (Byrnes, et al., 2015; Kim, et al., 2009). This section describes some of the recent advances in microfluidic SPrep for specific molecular diagnostic techniques like PCR, DNA sequencing etc. Table 3 summarizes some of these microfluidic platforms.

## Microfluidic Sample Preparation for PCR

PCR based methods have opened up a myriad of possibilities in diagnostics for pathogens and infectious diseases, in both clinical and environmental settings. There exists several commercial FDA approved PCR platforms (Priye & Ugaz, 2016), but most require either manual off-chip SPrep or separate automated SPrep systems involving bench top equipment like a centrifuge. Hence, these tests are not Clinical Laboratory Improvement Amendments (CLIA) waived or POC (Mitchell, et al., 2012). Hence, on-chip integration of SPrep will enable true sample-in, answer-out PCR platforms for POC use. They also provide advantages like lower reagent consumption, faster cycling times, lower cost per test, and automated processing requiring minimally trained personnel (Oblath, et al., 2014). Consistency of SPrep is known to affect results of digital, real-time PCR (RT-PCR) (Thompson, et al., 2014), so these advances are critical. Also microfluidic chips (µChip) can be disposable, eliminating contamination between samples, which is important since the high sensitivity provided by PCR poses issues due to NA contamination. This section focuses on recent advances in microfluidic devices with integrated SPrep-PCR and integrated SPrepPCR-detection capability. For a detailed review of microfluidic PCR and similar integrated systems prior to 2013, the reader can refer to (Park, et al., 2014).

Kim and co-workers developed a µChip that integrates solid-phase extraction and amplification of NAs into a single reaction chamber (Kim, et al., 2010). A nanoporous, aluminum oxide membrane (AOM) was employed for the solid-phase extraction of NAs. A µChip integrated DNA extraction using monolithic AOM and seven parallel reaction wells for real-time amplification of extracted DNAs. The system demonstrated the detection of bacterial pathogens in whole saliva sample (Oblath, et al., 2014). A disposable microfluidic chip with integrated solid phase extraction (SPE) for NA extraction and RT-PCR was used to amplify influenza A RNA in human nasopharyngeal aspirate and nasopharyngeal swab specimens (Mitchell, et al., 2012). However, the PCR products were detected off-chip by CE. In all of the devices mentioned above, sample lysis was performed off-chip.

A µChip integrating electrochemical cell lysis, PCR, CE based separation, and amperometric detection was reported for detection of pathogens (Jha, et al., 2012). However, the system displayed shortcomings associated with temperature control in PCR reactions (Adley, 2014). Czilwik and co-workers reported a centrifugal microfluidic based platform (LabDisk) utilizing pre-stored reagents, with integrated

DNA extraction, consensus multiplex PCR preamplification, and geometricallymultiplexed species-specific RT-PCR. The system was able to detect low concentrations of pathogens (2 CFU/200 μL) from serum samples (Czilwik, et al., 2015). The system requires serum separation from whole blood off-chip. However, serum or plasma separation has been demonstrated in microfluidic formats and could be integrated with the LabDisk system. Cai and co-workers demonstrated a completely integrated microfluidic device fabricated using "SlipChip" technologies for the detection of pathogens in biological samples (blood) (Cai, et al., 2014). The platform employs dielectrophoresis (DEP) for extraction, multiplex array PCR for amplification, and end-point fluorescence for the simultaneous detection of three different pathogens. However, the limit of detection of $10^3$ CFU/mL reported using the platform is low impeding its potential use in practical applications.

## Microfluidic SPrep for Isothermal Amplification

Isothermal amplification uses a single temperature as opposed to cycling between multiple temperatures as in the case of PCR. Since there is no thermal cycling involved, there is generally a reduced need for power, especially over long-term use, making it suitable for POC (Almassian & Nelson, 2013).

Huang and co-workers applied helicase-dependent isothermal amplification (HDA) for detection of *C. difficile* in stool samples. The electricity-free system consists of a μChip in a Styrofoam cup (the insulator), able to maintain its temperature at $65\pm2$ °C. SPrep employed a stand-alone pressure-driven "Portable System for Nucleic Acid Preparation" (SNAP), powered by a bicycle pump. It consisted of four subsystems: a sample input and mixer, a fluid buffering coil, an air pressure accumulator, and a sample extraction cartridge. The sample lysis and NA extraction was performed using the SNAP which was distinct from the amplification system, requiring manual transfer of extracted NA. The downstream detection of amplicons was also performed off-chip (Huang, et al., 2013). Hence, integration of sample lysis, NA extraction along with readout for amplicon detection would be necessary to achieve sample-in answer-out capability. A device consisting of flexible plastic substrate containing chambers, in a reel-to-reel cassette format, was used for the LAMP, and colorimetric detection. The system performs thermal shock lysis of hard to lyse Gram-positive bacteria on-chip, but fluid/reagent metering and mixing is done manually (pipetting) potentially limiting POC use (Safavieh, et al., 2014). Kim and co-workers developed a centrifugal microfluidic device integrating DNA extraction, isothermal recombinase polymerase amplification (RPA), and detection, onto a single disc (Figure 6). A laser diode was used for wireless control of valve actuation, cell lysis, and noncontact heating during RPA step. Immunomagnetic separation has proven to be an effective tool for preconcentration of pathogens from large volume samples containing potential interferents (Jayamohan, et al., 2015). However, for this device, the immunomagnetic extraction of pathogens was performed off-chip (Kim, et al., 2014). Similar centrifugal platforms have been applied to SPrep for RTPCR (viral detection) (Stumpf, et al.,

2016), LAMP (Sayad, et al., 2016) and digital PCR (Burger, et al., 2016; Schuler, et al., 2016). An eight-chamber LOC device integrating cell lysis, immunomagnetic bead based DNA extraction, LAMP, and fluorescence detection was reported (Sun, et al., 2015). The system was reported to have a true sample-in answer-out capability for the detection of *Salmonella*.

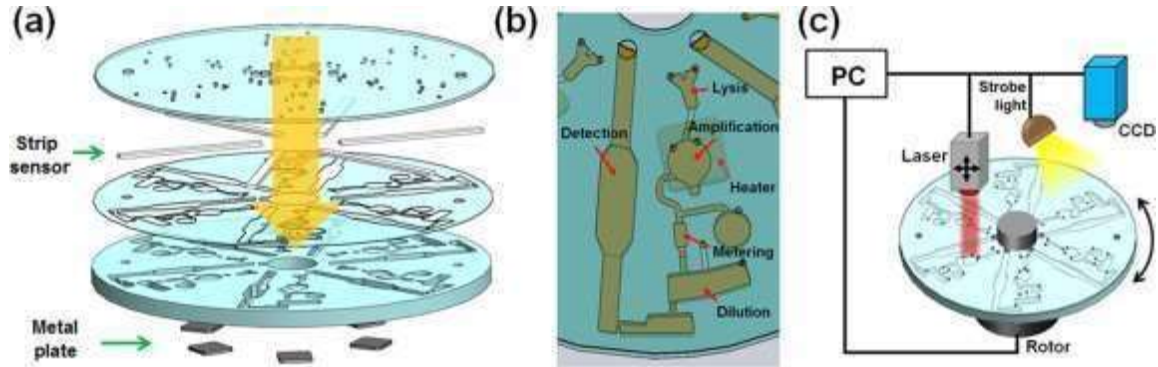

Figure 6. Lab-on-a-disc system for detection of food-borne pathogens. (a) The system consisted of two polycarbonate layers with integrated strip sensors and metal heaters. (b) The disc consists of chambers for cell lysis, isothermal amplification, metering, dilution, and detection. (c) Schematic of the setup showing computer controlled spinning motor, laser for the thermal actuation of ferrowax valves and isothermal DNA amplification, and a CCD camera and strobe light to visualize the rotating disc in real time. Reprinted with permission from (Kim, et al., 2014). Copyright 2014 American Chemical Society.

## Microfluidic SPrep for Sequencing

DNA sequencing refers to the process of determining the precise order of nucleotides within a DNA strand. Next-Generation DNA Sequencing (NGS) broadly refers to the recent advances in sequencing that has enabled high-throughput, inexpensive, rapid whole-genome sequencing (Metzker, 2010)(See also Chapter …). NGS has been widely applied to elucidate genetic information for applications like pathogen discovery and identification of genetic abnormalities associated with human disease (Kim, et al., 2013). Genome sequencing has come a long way since the conclusion of the Human Genome Project a decade back. Sequencing platforms have evolved from bulky systems (860 kg PacBios RSII) to relatively inexpensive, pocket-sized versions (Oxford MinION & SmidgION) (Erlich, 2015; Pennisi, 2016). Recent advances leveraging nanopores have the potential to democratize sequencing (Quick, et al., 2016). The cost of sequencing an individual full genome has plummeted from USD $2.7 billion (Human Genome Project) to the current $1000, outpacing even Moore's law (Hayden, 2014). Like in other areas of molecular diagnostics, automated DNA SPrep is one of the key challenges in achieving a small footprint, sample-in, data-out sequencing platform (Hayden, 2014; Coupland, 2010). For instance, advances in the area of automation of preparation methods for formatting sample DNA into sequencing ready libraries has lagged behind significant advances in NGS (Kim, et al.,

2013). However, recent advances have the potential to narrow the gap. This section will focus on recent published work on downstream microfluidic SPrep for genome sequencing (library preparation). We have published a detailed review of upstream microfluidic DNA SPrep techniques (cell lysis, DNA extraction), which the reader can refer to (Kim, et al., 2009).

Patel and co-workers developed a DMF platform as a fluid distribution hub (Figure 7). The platform enables the integration of multiple subsystem modules into an automated NGS library SPrep system. The central DMF hub is interfaced through novel capillary interconnects to external fluidic modules for highly repeatable transfer of liquid (Hanyoup, et al., 2011). The authors utilized a similar DMF platform for preparing sequencer-ready DNA libraries for analysis by Illumina MiSeq sequencer (Kim, et al., 2013). Cell lysis and DNA extraction steps were performed offchip using conventional laboratory methods. Tan and co-workers reported a novel microfluidic device, capable of performing an arbitrary number of serial reactionpurification steps on 16 independent samples. They applied the device to implement protocols for generating Next Generation DNA Sequencing libraries from bacterial and human genomic DNA samples. Similar DMF-based platforms (VolTRAX) are in the process of being commercialized for point-of-use automated sample preparation (Dodsworth, 2015; Oxford Nanopore, 2016).

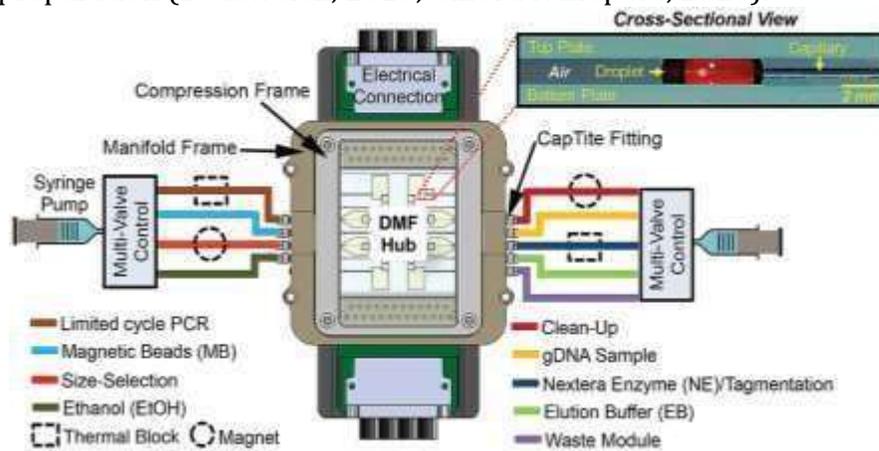

Figure 7. DMF system for preparing DNA libraries for sequencing. The system integrated multiple reagent and SPrep modules (depicted in different colors), magnets (for magnetic bead-based separation/cleanup) and thermal blocks (for thermal cycling) coupled to module tubing (for sample preparation), and multivalve syringe pumps (for liquid handling).

A single cell sequencing method (Drop-Seq), utilizing droplet microfluidics was proposed by Macosko et al (Macosko, et al., 2015). The system encapsulates one cell per emulsion droplet, lyses them, and uniquely barcodes the RNA of each cell using DNA-barcoded microgel beads. Hence, a number of conventional processing steps are compressed into a single step, creating a scalable method for *in-situ* library preparation (Erlich, 2015). Klein and co-workers reported a similar technique (inDrop), for barcoding the RNA from thousands of individual cells (Klein, et al., 2015). The technique was used to probe transcriptional variability in mouse embryonic stem cells. These methods had limitations seen in droplet microfluidics like variability in the number of cells per droplet (dictated by Poisson statistics). GnuBIO (a BioRad company) is commercializing a microfluidic-based system for genomic library preparation, integrated within their benchtop sequencer. The system promises to be a true "Sample In, Answer Out" DNA sequencing solution (Erlich, 2015; Business-Wire, 2014). Genetic analysis of minute amounts of DNA and RNA at the level of a single cell using NGS methods is increasingly being relied upon to understand biological complexity previously concealed when employing conventional techniques (Thompson, et al., 2014). These involve whole-genome amplification (WGA) or reverse transcription (RT) and WGA prior to NGS. SPrep improvements have helped improve the accuracy of the RT and preamplification steps. Still current RNA-sequencing methods cannot be considered as absolute counting technologies (Thompson, et al., 2014). Wu and co-workers performed a comparison of the sensitivity and reproducibility of single-cell whole transcriptome preparations. They reported less gene dropout and improved reproducibility and accuracy by performing RT and preamplification steps in microfluidic volumes of the C1 device (Fluidigm), rather than tube-based preparations (Thompson, et al., 2014).

In spite of significant advances in applying microfluidics in SPrep for sequencing, integrated sample-in sequence-out platforms are lacking. This is especially critical since real-time, portable sequencers are being developed (Quick, et al., 2016) and manual SPrep remains a critical bottleneck preventing their widespread use.

### Microfluidic SPrep in cytogenetics

Often in cytogenetic studies, the cell samples obtained are quite small and difficult to work with. In these cases, LOC devices have been designed that allow for culturing of many different kinds of cells. These devices often allow for greater efficiency when working with small volumes, as well as greater ease of operation and automation, which decreases the risk of human error (Tehranirokh, et al., 2013). SPrep is not limited to culturing, but also includes other processes that are necessary for FISH assays, karyotyping, or other cytogenetic testing. Shah and coworkers built a device, which integrates multiple stages of the process, allowing for initial culturing, arrest, and fixation of metaphase cells as well as having the ability to prepare metaphase chromosome spreads on glass slides for metaphase FISH analysis (Kwasny, et al., 2012). Creating the chromosome spreads has been described as more of an art than a science, but many devices (including the device built by Shah and co-workers) are being built which make the creation of these spreads more reliable and repeatable (Kwasny, et al., 2012; Kwasny, et al., 2014).

| Methods | Steps incorporated On-chip | Notes |
|---|---|---|
| PCR, RT-PCR (Kim, et al., 2010) | NA extraction (SPE) using nanoporous AOM & amplification | Inhibitory role of AOM during PCR, lower retention of RNA in AOM |
| RT-PCR (Oblath, et al., 2014) | NA extraction using nanoporous AOM & amplification | Off-chip thermal lysis, inhibitory effects of AOM minimized by adding BSA & additional *Taq* polymerase to the master mix |
| RT-PCR (Czilwik, et al., 2015) | Chemical lysis, NA extraction (silica coated magnetic beads), consensus pre-amplification & geometrically multiplexed RT-PCR | Demonstrated specific detection of the four model pathogens down to 3 CFUs in serum. The multiplex system incorporated pre-stored reagents, but PCR requires prior serum separation from whole blood, which might impede POC use. |
| Multiplex array PCR, DEP (Cai, et al., 2014) | Pathogen capture (DEP), thermal lysis & multiplex array PCR | Simultaneous detection of three pathogens in 3 hours. Preloaded PCR reagents and sample-in answer-out capability for potential POC use. But the system relied on a conventional thermal cycler, power supply (for DEP) and fluorescence microscope, potentially impacting its portability for POC use. |
| HDA (Huang, et al., 2013) | Chemical lysis, NA extraction (SPE) & HDA amplification | The lysis and NA extraction was performed on a based system (SNAP) distinct from the NA amplification. The electricity-free NA amplification system consisted of a cyclic olefin polymer μChip placed inside a Styrofoam cup with commercially available toe warmers acting as heaters. The extracted NA was manually transferred over to the μChip for amplification. |
| RPA (Kim, et al., 2014) | Thermal lysis, RPA based amplification & visual detection using lateral flow strips | Centrifugal platform for food borne pathogen detection. Laser diode used for actuation of ferrowax valves, thermal lysis and amplification. |

| Real-time LAMP (Sun, et al., 2015) | Chemical lysis, NA extraction (magnetic beads), isothermal amplification | Integration of SPrep to LAMP based detection on same eight chamber thermoplastic chip. |
|---|---|---|
| Convective PCR (Priye, et al., 2016) | Chemical lysis, NA extraction (SPE), PCR amplification, fluorescence detection | Portable convective thermocycler loaded on a drone, with SPrep leveraging drone's motors. Time-resolved fluorescence detection and quantification using integrated smartphone camera. Sample lysis and loading is performed manually utilizing the platform. |
| NGS Library prep (Kim, et al., 2013) | NGS library prep | DMF platform to prepare NGS libraries from few nanograms of genomic DNA. DNA extraction and purification performed off-chip. |

Table 3. Microfluidic SPrep platforms used in NA amplification and NGS, with onchip SPrep steps listed.

### Microfluidics in Cell sorting

Cell sorting can be performed using both electrokinetically and hydrodynamically driven mechanisms in microfluidic devices (Paegel, et al., 2003). Fu and co-workers demonstrated single-cell handling feasibility (Fu, et al., 2002). Grove and co-workers present a hybrid glass-PDMS microfluidic device with elastomer valves and pumps, which provided reliable fluid control on chip (Grover, et al., 2003). In 2010, Gagnon and co-workers developed a closed-loop microfluidic device for yeast cell separation using AC electrokinetic components (Gagnon, et al., 2010). In 2011, a microfluidic device used for high-efficiency circulating tumor cell selection was presented (Dharmasiri, et al., 2011). Karimi and co-workers reported the cell focusing and sorting using hydrodynamic mechanisms (Karimi, et al., 2013). More recently, Song and co-workers developed an electrokinetic microfluidic device for cell sorting (Song, et al., 2015).

### Future of Microfluidics for Medical Diagnostics

The medical diagnostics field is rapidly being transformed by the introduction and optimization of microfluidic devices. Because of the obvious size match between microfluidics and most biological processes at the cellular and subcellular level, the use of microfluidics will only naturally continue to be applied to medical diagnostics. As complexity is much more readily introduced at the microscale than in traditional formats, it is anticipated that more and more complex microfluidic devices able to perform multiple diagnostic processes at the same time will be developed. Already there are systems that can perform detection of multiple diseases or pathogens simultaneously. These devices currently rely on using the same methods on multiple targets. As sample preparation techniques improve, though, it is anticipated that multiple processes will be possible on one chip, allowing detection of DNA, proteins, chemicals, and other biomolecules on one device. As multiplexed protein detection joins multiplexed DNA detection and analysis, small-scale, portable, microfluidicbased instruments will be able to perform massively parallel analysis simultaneously at lower and lower costs, making the need for individual tests less relevant, especially when physicians are attempting to make a diagnosis with symptoms suggesting any of several possible conditions. Thus, the dream of a single instrument to perform nearly any molecular diagnostic procedure can be seen now and will likely occur in the next 10-20 years

Even further along, though, may be the opportunity for combinations of microfluidics and nucleic acid sequencing technologies. Sequencing is becoming ubiquitous with prices falling rapidly. As sequencing becomes a commodity, can be completed more rapidly, and the instruments are further miniaturized, sample preparation using microfluidic instruments will become the limiting factor in developing "near universal" molecular diagnostic tools, at least for anything that can be diagnosed using a nucleic acid sequence. Of course, the biological understanding and computer software will need to keep up, but it is clear that both will move quickly if the opportunity is available. Thus, universal sample preparation instruments for highthroughput sequencing may be one of the major challenges for microfluidics in the next decade and for the foreseeable future of molecular diagnostics. Rapid development in this area will only be possible as more robust, generic, and complex microfluidics allow, making this a potent area for high-impact research activities.

## Bibliography

Abate, A. R. et al., 2013. DNA sequence analysis with droplet-based microfluidics. *Lab on a Chip,* 13(24), pp. 4864-4869.

Adley, C. C., 2014. Past, present and future of sensors in food production. 3(3), pp. 491-510.

Aebersold, R. & Mann, M., 2003. Mass spectrometry-based proteomics. *Nature,* 422(6928), pp. 198-207.

Ahlford, A. et al., 2011. *A microfluidic platform for personalized cancer diagnostics by padlock probe ligation and circle-to-circle amplification ..* Seattle, s.n., pp. 61-63.

Almassian, D. R. C. L. M. & Nelson, W. M., 2013. Portable nucleic acid thermocyclers. *Chemical Society reviews,* 42(22), pp. 8769-8798.

Almoguera, C. et al., 1988. Most human carcinomas of the exocrine pancreas contain mutant cK-ras genes. *Cell,* 53(4), pp. 549-554.

Asiello, P. J. & Baeumner, A. J., 2011. Miniaturized isothermal nucleic acid amplification, a review. *Lab on a Chip,* 11(8), pp. 1420-1430.

Athamanolap, P., Shin, D. J. & Wang, T.-H., 2014. Droplet array platform for highresolution melt analysis of DNA methylation density. *Journal of laboratory automation,* 19(3), pp. 304-312.

Baker, C. A. & Roper, M. G., 2012. Online coupling of digital microfluidic devices with mass spectrometry detection using an eductor with electrospray ionization. *Analytical chemistry,* 84(6), pp. 2955-2960.

Becker, H., 2010. Mind the gap!. *Lab on a Chip,* pp. 271-273.

Blazej, R. G., Kumaresan, P. & Mathies, R. A., 2006. Microfabricated bioprocessor for integrated nanoliter-scale Sanger DNA sequencing. *Proceedings of the National Academy of Sciences,* 103(19), pp. 7240-7245.

Brehm-Stecher, B., Young, C., Jaykus, L.-A. & Tortorello, M. L., 2009. Sample Preparation: The Forgotten Beginning. *Journal of Food Protection,* 72(8), pp. 17741789.

Burger, R., Amato, L. & Boisen, A., 2016. Detection methods for centrifugal microfluidic platforms.". *Biosensors and Bioelectronics,* Volume 76, pp. 54-67.

Business-Wire, 2014. *Bio-Rad Acquires GnuBIO, Developer of Droplet-Based DNA Sequencing Technology.* [Online]
Available at: http://gnubio.com/news/bio-rad-acquires-gnubio-developerofdroplet-based-dna-sequencing-technology-2/ [Accessed 12 September 2015].

Byrnes, S. A. et al., 2015. One-step purification and concentration of DNA in porous membranes for point-of-care applications. *Lab on a Chip.*

Cai, D. et al., 2014. An integrated microfluidic device utilizing dielectrophoresis and multiplex array PCR for point-of-care detection of pathogens. *Lab on a Chip,* 14(20), pp. 3917-3924.

Cao, W. et al., 2015. Automated Microfluidic Platform for Serial Polymerase Chain Reaction and High-Resolution Melting Analysis. *Journal of laboratory automation,* p. 2211068215579015.

Chang, C.-M.et al., 2013. Nucleic acid amplification using microfluidic systems. *Lab on a Chip,* 13(7), pp. 1225-1242.

Chao, T. & Hansmeier, N., 2013. Microfluidic devices for high-throughput proteome analyses. *Proteomics,* 13(3-4), pp. 467-479.

Chatterjee, D. et al., 2010. Integration of protein processing steps on a droplet microfluidics platform for MALDI-MS analysis. *Analytical chemistry,* 82(5), pp. 20952101.

Chin, C. D. et al., 2011. Microfluidics-based diagnostics of infectious diseases in the developing world. *Nature medicine,* 17(8), pp. 1015-1019.

Chin, C. D., Linder, V. & Sia, S. K., 2012. Commercialization of microfluidic point-ofcare diagnostic devices. *Lab on a Chip,* 12(12), pp. 2118-2134.

Coupland, P., 2010. Microfluidics for the upstream pipeline of DNA sequencing–a worthy application?. *Lab on a Chip,* 10(5), pp. 544-547.

Craw, P. & Balachandran, W., 2012. Isothermal nucleic acid amplification technologies for point-of-care diagnostics: a critical review. *Lab on a Chip,* 12(14), pp. 2469-2486.

Crews, N., Wittwer, C., Palais, R. & Gale, B., 2008. Product differentiation during continuous-flow thermal gradient PCR. *Lab on a Chip,* 8(6), pp. 919-924. Crews, N. et al., 2009. Spatial DNA melting analysis for genotyping and variant scanning. *Analytical chemistry,* 81(6), pp. 2053-2058.

Czilwik, G. et al., 2015. Rapid and fully automated bacterial pathogen detection on a centrifugal-microfluidic LabDisk using highly sensitive nested PCR with integrated sample preparation. *Lab on a Chip,* 15(18), pp. 3749-3759.

Derveaux, S. et al., 2008. Synergism between particle-based multiplexing and microfluidics technologies may bring diagnostics closer to the patient. *Analytical and bioanalytical chemistry,* 391(7), pp. 2453-2467.

DeVoe, D. L. & Lee, C. S., 2006. Microfluidic technologies for MALDI-MS in proteomics. *Electrophoresis,* 27(18), pp. 3559-3568.

Dharmasiri, U. et al., 2011. High-throughput selection, enumeration, electrokinetic manipulation, and molecular profiling of low-abundance circulating tumor cells using a microfluidic system. *Analytical chemistry,* 83(6), pp. 2301-2309.

Dodsworth, S., 2015. Genome skimming for next-generation biodiversity analysis. *Trends in plant science,* 20(9), pp. 525-527.

Domon, B. & Aebersold, R., 2006. Mass spectrometry and protein analysis. *Science,* 312(5771), pp. 212-217.

Duffy, D. C. J., McDonald, C., Schueller, O. J. & Whitesides, G. M., 1998. Rapid prototyping of microfluidic systems in poly (dimethylsiloxane). *Analytical chemistry,* pp. 4974-4984.

Erlich, Y., 2015. A vision for ubiquitous sequencing. *bioRxiv,* p. 019018. Farrar, J. S. & Wittwer, C. T., 2015. Extreme PCR: Efficient and Specific DNA Amplification in 15-60 Seconds. *Clinical Chemisty,* Volume 61, pp. 145-153.

Fredrickson, C. K. & Fan, Z. H., 2004. Macro-to-micro interfaces for microfluidic devices. *Lab on a Chip,* pp. 526-533.

Fu, A. Y. et al., 2002. An integrated microfabricated cell sorter. *Analytical Chemistry,* 74(11), pp. 2451-2457.

Gagnon, Z., Mazur, J. & Chang, H.-C., 2010. Integrated AC electrokinetic cell separation in a closed-loop device. *Lab on a Chip,* 10(6), pp. 718-726. Gao, D., Liu, H., Jiang, Y. & Lin, J.-M., 2013. Recent advances in microfluidics combined with mass spectrometry: technologies and applications. *Lab on a Chip,* 13(17), pp. 3309-3322.

Gonzalez de Castro, D., Clarke, P. A., Al-Lazikani, B. & Workman, P., 2013. Personalized cancer medicine: molecular diagnostics, predictive biomarkers, and drug resistance. *Clinical Pharmacology & Therapeutics,* 93(3), pp. 252-259. Grover,

W. H. et al., 2003. Monolithic membrane valves and diaphragm pumps for practical large-scale integration into glass microfluidic devices. *Sensors and Actuators B: Chemical,* 89(3), pp. 315-323.

Hadd, A. G., Goard, M. P., Rank, D. R. & Jovanovich, S. B., 2000. Sub-microliter DNA sequencing for capillary array electrophoresis. *Journal of Chromatography A,* 894(1), pp. 191-201.

Han, X. & Gross, R. W., 2005. Shotgun lipidomics: electrospray ionization mass spectrometric analysis and quantitation of cellular lipidomes directly from crude extracts of biological samples. *Mass spectrometry reviews,* 24(3), pp. 367-412.

Hanyoup, K. et al., 2011. Automated Digital Microfluidic Sample Preparation for Next-Generation DNA Sequencing. *Journal of the Association for Laboratory Automation,* 16(6), pp. 405-414.

Hardouin, J., 2007. Protein sequence information by matrix-assisted laser desorption/ionization in-source decay mass spectrometry. *Mass spectrometry reviews,* 26(5), pp. 672-682.

Hauck, T. S., Giri, S., Gao, Y. & Chan, W. C., 2010. Nanotechnology diagnostics for infectious diseases prevalent in developing countries. *Advanced drug delivery reviews,* 62(4), pp. 438-448.

Hayden, E. C., 2014. Technology: The $1,000 genome. *Nature News,* 507(7492), p. 294–295.

He, M. & Herr, A. E., 2010. Automated microfluidic protein immunoblotting. *Nature protocols,* 5(11), pp. 1844-1856.

He, M., Novak, J., Julian, B. A. & Herr, A. E., 2011. Membrane-assisted online renaturation for automated microfluidic lectin blotting. *Journal of the American Chemical Society,* 133(49), pp. 19610-19613.

He, Y., Pang, H.-M. & Yeung, E. S., 2000. Integrated electroosmotically-driven on-line sample purification system for nanoliter DNA sequencing by capillary electrophoresis. *Journal of Chromatography A,* 894(1), pp. 179-190.

Hindson, B. J. N. K. D. et al., 2011. High-throughput droplet digital PCR system for absolute quantitation of DNA copy number. *Analytical chemistry,* 83(22), pp. 86048610.

Ho, J. et al., 2011. Paper-based microfluidic surface acoustic wave sample delivery and ionization source for rapid and sensitive ambient mass spectrometry. *Analytical chemistry,* 83(9), pp. 3260-3266.

Hsieh, A. T.-H., Pan, P. J.-H. & Lee, A. P., 2009. Rapid label-free DNA analysis in picoliter microfluidic droplets using FRET probes. *Microfluidics and nanofluidics,* 6(3), pp. 391-401.

Huang, S. et al., 2013. Low cost extraction and isothermal amplification of DNA for infectious diarrhea diagnosis. *PloS one,* 8(3).

Hughes, A. J. & Herr, A. E., 2012. Microfluidic western blotting. *Proceedings of the National Academy of Sciences,* 109(52), pp. 21450-21455.

Jacobs, J. M. et al., 2005. Utilizing human blood plasma for proteomic biomarker discovery. *Journal of proteome research,* 4(4), pp. 1073-1085.

Jansson, R., 2007. *Development of a solid-phase padlock probe technology using microfluidics.* Uppsala: Uppsala University School of Engineering.

Jarvius, J. et al., 2006. Digital quantification using amplified single-molecule detection. *Nature methods,* 3(9), pp. 725-727.

Jayamohan, H. et al., 2015. Highly Sensitive Bacteria Quantification Using Immunomagnetic Separation and Electrochemical Detection of Guanine-Labeled Secondary Beads. *Sensors,* 15(5), pp. 12034-12052.

Jayamohan, H., Sant, H. J. & Gale, B. K., 2013. Applications of microfluidics for molecular diagnostics. In: *Microfluidic Diagnostics.* s.l.:Springer, pp. 305-334.

Jenkins, G. & Mansfield, C. D., 2013. *Microfluidic Diagnostics: Methods and Protocols.* s.l.:Humana Press.

Jha, S. K. et al., 2012. An integrated PCR microfluidic chip incorporating aseptic electrochemical cell lysis and capillary electrophoresis amperometric DNA detection for rapid and quantitative genetic analysis. *Lab on a Chip,* p. 4455–4464. Ji, J. et al., 2012. Proteolysis in microfluidic droplets: an approach to interface protein separation and peptide mass spectrometry. *Lab on a Chip,* 12(15), pp. 26252629.

Jin, S. G. J. A. a. R. T. K., 2013. Western blotting using microchip electrophoresis interfaced to a protein capture membrane. *Analytical chemistry,* 85(12), pp. 60736079.

Jin, S. & Kennedy, R. T., 2015. New developments in Western blot technology. *Chinese Chemical Letters,* 26(4), pp. 416-418.

Kalra, H. et al., 2013. Comparative proteomics evaluation of plasma exosome isolation techniques and assessment of the stability of exosomes in normal human blood plasma. *Proteomics,* 13(22), pp. 3354-3364.

Karimi, A., Yazdi, S. & Ardekani, A., 2013. Hydrodynamic mechanisms of cell and particle trapping in microfluidics. *Biomicrofluidics,* 7(2), p. 021501.

Karns, K. & Herr, A. E., 2011. Human tear protein analysis enabled by an alkaline microfluidic homogeneous immunoassay. *Analytical chemistry,* 83(21), pp. 81158122.

Khandurina, J., Chován, T. & Guttman, A., 2002. Micropreparative fraction collection in microfluidic devices. *Analytical chemistry,* 74(7), pp. 1737-1740. Kim, D. et al., 2012. Electrostatic protein immobilization using charged polyacrylamide gels and cationic detergent microfluidic Western blotting. *Analytical chemistry,* 84(5), pp. 2533-2540.

Kim, H. et al., 2013. A Microfluidic DNA Library Preparation Platform for NextGeneration Sequencing. *PLOS one,* 8(7), p. e68988.

Kim, J., Johnson, M., Hill, P. & Gale, B. K., 2009. Microfluidic sample preparation: cell lysis and nucleic acid purification. *Integrative Biology,* pp. 574-586.

Kim, J. et al., 2010. A PCR reactor with an integrated alumina membrane for nucleic acid isolation. *Analyst,* 135(9), pp. 2408-2414.

Kim, T.-H., Park, J., Kim, C.-J. & Cho, Y.-K., 2014. Fully integrated lab-on-a-disc for nucleic acid analysis of food-borne pathogens. *Analytical chemistry,* 86(8), pp. 38413848.

Klein, A. M. et al., 2015. Droplet Barcoding for Single-Cell Transcriptomics Applied to Embryonic Stem Cells. *Cell,* 161(5), pp. 1187-1201.

Klostranec, J. M. et al., 2007. Convergence of quantum dot barcodes with microfluidics and signal processing for multiplexed high-throughput infectious disease diagnostics. *Nano Letters,* 7(9), pp. 2812-2818.

Konry, T. et al., 2011. Ultrasensitive Detection of Low-Abundance Surface-Marker Protein Using Isothermal Rolling Circle Amplification in a Microfluidic Nanoliter Platform. *Small,* 7(3), pp. 395-400.

Kühnemund, M., Witters, D., Nilsson, M. & Lammertyn, J., 2014. Circle-to-circle amplification on a digital microfluidic chip for amplified single molecule detection. *Lab on a Chip,* 14(16), pp. 2983-2992.

Kuroda, A. et al., 2014. Microfluidics-based in situ padlock/rolling circle amplification system for counting single DNA molecules in a cell. *Analytical Sciences,* 30(12), pp. 1107-1112.

Küster, S. K. et al., 2013. Interfacing droplet microfluidics with matrix-assisted laser desorption/ionization mass spectrometry: label-free content analysis of single droplets. *Analytical chemistry,* 85(3), pp. 1285-1289.

Kwasny, D. et al., 2014. A Semi-Closed Device for Chromosome Spreading for Cytogenetic Analysis. *Micromachines,* 5(2), pp. 158-170.

Kwasny, D. et al., 2012. Microtechnologies Enable Cytogenetics. In: P. Tirunilai, ed. *Recent Trends in Cytogenetic Studies-Methodologies and Applications .* Rijeka: INTECH Open Access Publisher, pp. 20-46.

Lagally, E. T., Simpson, P. C. & Mathies, R. A., 2000. Monolithic integrated microfluidic DNA amplification and capillary electrophoresis analysis system. *Sensors and Actuators B: Chemical,* 63(3), pp. 138-146.

Lapierre, F. et al., 2011. High sensitive matrix-free mass spectrometry analysis of peptides using silicon nanowires-based digital microfluidic device. *Lab on a Chip,* 11(9), pp. 1620-1628.

Lazar, I. M. & Kabulski, J. L., 2013. Microfluidic LC device with orthogonal sample extraction for on-chip MALDI-MS detection. *Lab on a Chip,* 13(11), pp. 2055-2065.

Lazar, I. M., Trisiripisal, P. & Sarvaiya, H. A., 2006. Microfluidic liquid chromatography system for proteomic applications and biomarker screening. *Analytical chemistry,* 78(15), pp. 5513-5524.

Lee, A., 2013. The third decade of microfluidics. *Lab on a Chip,* pp. 1660-1661. Lee, J., Soper, S. A. & Murray, K. K., 2009. Microfluidics with MALDI analysis for proteomics-A review. *Analytica chimica acta,* 649(2), pp. 180-190.

Lee, W. & Fan, X., 2012. Intracavity DNA melting analysis with optofluidic lasers. *Analytical chemistry,* 84(21), pp. 9558-9563.

Liu, S., Shi, Y., Ja, W. W. & Mathies, R. A., 1999. Optimization of high-speed DNA sequencing on microfabricated capillary electrophoresis channels. *Analytical Chemistry,* 71(3), pp. 566-573.

Li, Z. et al., 2010. Rapid and sensitive detection of protein biomarker using a portable fluorescence biosensor based on quantum dots and a lateral flow test strip. *Analytical chemistry,* 82(16), pp. 7008-7014.

Lochovsky, C., Yasotharan, S. & Günther, A., 2012. Bubbles no more: in-plane trapping and removal of bubbles in microfluidic devices. *Lab on a Chip,* 12(3), pp. 595-601.

Macosko, E. Z. et al., 2015. Highly Parallel Genome-wide Expression Profiling of Individual Cells Using Nanoliter Droplets. *Cell,* 161(5), pp. 1202-1214. Mahmoudian, L. et al., 2008. Rolling circle amplification and circle-to-circle amplification of a specific gene integrated with electrophoretic analysis on a single chip. *Analytical chemistry,* 80(7), pp. 2483-2490.

Mani, V. et al., 2009. Ultrasensitive immunosensor for cancer biomarker proteins using gold nanoparticle film electrodes and multienzyme-particle amplification. *ACS nano,* 3(3), pp. 585-594.

Manz, A. N. G. a. H. M. W., 1990. Miniaturized total chemical analysis systems: a novel concept for chemical sensing. *Sensors and actuators B: Chemical,* pp. 244-248.. Mao, X. & Huang, T. J., 2012. Microfluidic diagnostics for the developing world. *Lab on a Chip,* 12(8), pp. 1412-1416.

Mariella Jr, R., 2008. Sample preparation: the weak link in microfluidics-based biodetection. *Biomed Microdevices.*

Melin, J. et al., 2005. Melin, Jonas, Henrik Johansson, Ola Söderberg, Fredrik Nikolajeff, Ulf Landegren, Mats Nilsson, and Jonas Jarvius. "Thermoplastic microfluidic platform for single-molecule detection, cell culture, and actuation. *Analytical chemistry,* 77(22), pp. 7122-7130.

Mery, E. et al., 2008. A silicon microfluidic chip integrating an ordered micropillar array separation column and a nano-electrospray emitter for LC/MS analysis of peptides. *Sensors and Actuators B: Chemical,* 134(2), pp. 438-446.

Metzker, M. L., 2010. Sequencing technologies-the next generation. *Nature reviews genetics,* 11(1), pp. 31-46.

Mezger, A. et al., 2014. Detection of Rotavirus Using Padlock Probes and Rolling Circle Amplification. *PloS one,* 9(11), p. e111874.

Mitchell, P., Feldman, J., Pollock, N. R. & Klapperich, C. M., 2012. Microfluidic chip for molecular amplification of influenza A RNA in human respiratory specimens. *PLoS One,* 7(3), p. e33176.

Mitchell, P. S. et al., 2008. Circulating microRNAs as stable blood-based markers for cancer detection. *Proceedings of the National Academy of Sciences,* 105(30), pp. 10513-10518.

Neuzil, P. et al., 2006. Ultra fast miniaturized real-time PCR: 40 cycles in less than six minutes. *Nucleic acids research,* 34(11), pp. e77-e77.

Ng, A. H., Uddayasankar, U. & Wheeler, A. R., 2010. Immunoassays in microfluidic systems. *Analytical and bioanalytical chemistry,* 397(3), pp. 991-1007.
Northen, T. R. et al., 2007. Clathrate nanostructures for mass spectrometry. *Nature,* 449(7165), pp. 1033-1036.
Oblath, E. A., Henley, W. H., Alarie, J. P. & Ramsey, J. M., 2013. A microfluidic chip integrating DNA extraction and real-time PCR for the detection of bacteria in saliva. *Lab on a Chip,* 13(7), pp. 1325-1332.
Oblath, E. A., Henley, W. H., Alarie, J. P. & Ramsey, J. M., 2014. A microfluidic chip integrating DNA extraction and real-time PCR for the detection of bacteria in saliva. *Lab on a Chip,* 13(7), pp. 1325-1332.
Oxford Nanopore, 2016. *VolTRAX: rapid, programmable, portable, disposable sample processor.* [Online]
Available at: https://publications.nanoporetech.com/2016/05/26/voltraxrapidprogrammable-portable-disposable-sample-processor/
Paegel, B. M., Blazej, R. G. & Mathies, R. A., 2003. Microfluidic devices for DNA sequencing: sample preparation and electrophoretic analysis. *Current opinion in biotechnology,* 14(1), pp. 42-50.
Paegel, B. M. et al., 2002. High throughput DNA sequencing with a microfabricated 96-lane capillary array electrophoresis bioprocessor. *Proceedings of the National Academy of Sciences,* 99(2), pp. 574-579.
Paliwal, A., Vaissière, T. & Herceg, Z., 2010. Quantitative detection of DNA methylation states in minute amounts of DNA from body fluids. *Methods,* 52(3), pp. 242-247.
Park, B. H., Kim, Y. T., Jung, J. H. & Seo, T. S., 2014. Integration of sample pretreatment, μPCR, and detection for a total genetic analysis microsystem. *Microchimica Acta,* 181(13-14), pp. 1655-1668.
Pekin, D. et al., 2011. Quantitative and sensitive detection of rare mutations using droplet-based microfluidics. *Lab on a Chip,* 11(13), pp. 2156-2166.
Pennisi, E., 2016. Pocket DNA sequencers make real-time diagnostics a reality. *Science,* 351(6275), pp. 800-801.
Perez-Toralla, K. et al., 2015. FISH in chips: turning microfluidic fluorescence in situ hybridization into a quantitative and clinically reliable molecular diagnosis tool. *Lab on a Chip,* 15(3), pp. 811-822.
Pješčić, I. & Crews, N., 2012. Genotyping from saliva with a one-step microdevice. *Lab on a Chip,* 12(14), pp. 2514-2519.
Pješčić, I. et al., 2011. Real-time damage monitoring of irradiated DNA. *Integrative Biology,* 3(9), pp. 937-947.
Pješčić, I., Tranter, C., Hindmarsh, P. L. & Crews, N. D., 2010. Glass-composite prototyping for flow PCR with in situ DNA analysis. *Biomedical microdevices,* 12(2), pp. 333-343.

Prakash, R. et al., 2014. Multiplex, Quantitative, Reverse Transcription PCR Detection of Influenza Viruses Using Droplet Microfluidic Technology. *Micromachines,* 6(1), pp. 63-79.

Priye, A. & Ugaz, V. M., 2016. Convective PCR Thermocycling with SmartphoneBased Detection: A Versatile Platform for Rapid, Inexpensive, and Robust Mobile Diagnostics. In: C. Lu & S. S. Verbridge, eds. *Microfluidic Methods for Molecular Biology.* s.l.:Springer International Publishing, pp. 55-69.

Priye, A. et al., 2016. Lab-on-a-drone: Toward pinpoint deployment of smartphoneenabled nucleic acid-based diagnostics for mobile health care. *Analytical chemistry,* 88(9), pp. 4651-4660.

Quick, J. et al., 2016. Real-time, portable genome sequencing for Ebola surveillance. *Nature,* 530(7589), pp. 228-232.

Ro, K. W., Liu, J. & Knapp, D. R., 2006. Plastic microchip liquid chromatographymatrix-assisted laser desorption/ionization mass spectrometry using monolithic columns. *Journal of chromatography A,* 1111(1), pp. 40-47.

Romanov, V. S. N. D. A. R. M. D. W. G. B. K. G. a. B. D. B., 2014. A critical comparison of protein microarray fabrication technologies. *Analyst,* 139(6), pp. 1303-1326.

Safavieh, M. et al., 2014. A simple cassette as point-of-care diagnostic device for naked-eye colorimetric bacteria detection. *Analyst,* 139(2), pp. 482-487. Sainiemi, L. et al., 2012. A microfabricated micropillar liquid chromatographic chip monolithically integrated with an electrospray ionization tip. *Lab on a Chip,* 12(2), pp. 325-332.

Samuel, R. et al., 2016. *40 Cycle PCR Using Human Genomic DNA In Less Than 1 Minute.* Dublin, Ireland, Proceedings of MicroTAS 2016.

Sanvicens, N., Pastells, C., Pascual, N. & Marco, M.-P., 2009. Nanoparticle-based biosensors for detection of pathogenic bacteria. *TrAC Trends in Analytical Chemistry,* pp. 1243-1252.

Sato, K. A. T. B. R. K. M. K. S. Y. T. J. J. M. N. a. T. K., 2010. Microbead-based rolling circle amplification in a microchip for sensitive DNA detection. *Lab on a Chip,* 10(10), pp. 1262-1266.

Sayad, A. A. et al., 2016. A microfluidic lab-on-a-disc integrated loop mediated isothermal amplification for foodborne pathogen detection. *Sensors and Actuators B: Chemical ,* Volume 227, pp. 600-609.

Schoenitz, M., Grundemann, L., Augustin, W. & Scholl, S., 2015. Fouling in microstructured devices: a review. *Chem. Commun.,* p. 8213.

Schuler, F. et al., 2016. Digital droplet PCR on disk. *Lab on a Chip,* 16(1), pp. 208-216.

Shamsi, M. H., Choi, K., Ng, A. H. & Wheeler, A. R., 2014. A digital microfluidic electrochemical immunoassay. *Lab on a Chip,* 14(3), pp. 547-554.

Shih, S. C. et al., 2012. Dried blood spot analysis by digital microfluidics coupled to nanoelectrospray ionization mass spectrometry. *Analytical chemistry,* 84(8), pp. 3731-3738.

Song, Y. et al., 2015. Size-based cell sorting with a resistive pulse sensor and an electromagnetic pump in a microfluidic chip. *Electrophoresis,* 36(3), pp. 398-404.

Srbek, J. J. E. U. E. K. K. T. v. d. G. a. P. C., 2007. Chip-based nano-LC-MS/MS identification of proteins in complex biological samples using a novel polymer microfluidic device. *Journal of separation science,* 30(13), pp. 2046-2052.

Stumpf, F. et al., 2016. LabDisk with complete reagent prestorage for sample-toanswer nucleic acid based detection of respiratory pathogens verified with influenza A H3N2 virus. *Lab on a Chip,* 16(1), pp. 199-207.

Sundberg, S. O. et al., 2014. Quasi-digital PCR: Enrichment and quantification of rare DNA variants. *Biomedical microdevices,* 16(4), pp. 639-644.

Sun, Y. et al., 2015. A lab-on-a-chip system with integrated sample preparation and loop-mediated isothermal amplification for rapid and quantitative detection of Salmonella spp. in food samples. *Lab on a Chip,* 15(8), pp. 1898-1904.

Tanaka, Y. et al., 2011. Single-molecule DNA patterning and detection by padlock probing and rolling circle amplification in microchannels for analysis of small sample volumes. *Analytical chemistry,* 83(9), pp. 3352-3357.

Tehranirokh, M., Kouzani, A. Z., Francis, P. S. & Kanwar, J. R., 2013. Microfluidic devices for cell cultivation and proliferation. *Biomicrofluidics,* 7(5), p. 051502.

Teh, S.-Y., Lin, R., Hung, L.-H. & Lee, A. P., 2008. Droplet microfluidics. *Lab on a Chip,* 8(2), pp. 198-220.

Thompson, A. M. et al., 2014. Microfluidics for single-cell genetic analysis. *Lab on a Chip,* 14(7), pp. 3135-3142.

Thuy, T. T., Inganäs, M., Ekstrand, G. & Thorsén, G., 2010. Parallel sample preparation of proteins, from crude samples to crystals ready for MALDI-MS, in an integrated microfluidic system. *Journal of Chromatography B,* 878(28), pp. 28032810.

Tian, H., Hühmer, A. F. & Landers, J. P., 2000. Evaluation of silica resins for direct and efficient extraction of DNA from complex biological matrices in a miniaturized format. *Analytical biochemistry,* 283(2), pp. 175-191.

Tröger, V., K., N., Gärtig, C. & Kuhlmeier, D., 2015. Isothermal Amplification and Quantification of Nucleic Acids and its Use in Microsystems. *Journal of Nanomedicine & Nanotechnology,* 6(282), p. 2.

Unger, M. A. et al., 2000. Monolithic microfabricated valves and pumps by multilayer soft lithography. *Science,* pp. 113-116.

Vergauwe, N. et al., 2011. A versatile electrowetting-based digital microfluidic platform for quantitative homogeneous and heterogeneous bio-assays. *Journal of Micromechanics and Microengineering,* 21(5), p. 054026.

Volpatti, L. R. & Yetisen, A. K., 2014. Commercialization of microfluidic devices. *Trends in biotechnology,* 32(7), pp. 347-350.

Weisenberger, D. J. et al., 2008. DNA methylation analysis by digital bisulfite genomic sequencing and digital MethyLight. *Nucleic acids research,* 36(14), pp. 4689-4698.

Wheeler, A. R. et al., 2005. Digital microfluidics with in-line sample purification for proteomics analyses with MALDI-MS. *Analytical Chemistry,* 77(2), pp. 534-540.
Whitesides, G. M., 2006. The origins and the future of microfluidics. *Nature,* pp. 368373.
Woolley, A. T. & Mathies, R. A., 1995. Ultra-high-speed DNA sequencing using capillary electrophoresis chips. 67(20), pp. 3676-3680.
Xie, J. et al., 2005. Microfluidic platform for liquid chromatography-tandem mass spectrometry analyses of complex peptide mixtures. *Analytical chemistry,* 77(21), pp. 6947-6953.
Xue, G., Pang, H.-M. & Yeung, E. S., 2001. On-line nanoliter cycle sequencing reaction with capillary zone electrophoresis purification for DNA sequencing. *Journal of Chromatography A,* 914(1), pp. 245-256.
Xu, Z.-R.et al., 2010. A miniaturized spatial temperature gradient capillary electrophoresis system with radiative heating and automated sample introduction for DNA mutation detection. *Electrophoresis,* 31(18), pp. 3137-3143.
Yetisen, A. K. & Volpatti, L. R., 2014. Patent protection and licensing in microfluidics. *Lab on a Chip,* pp. 2217-2225.
Yin, H. et al., 2005. Microfluidic chip for peptide analysis with an integrated HPLC column, sample enrichment column, and nanoelectrospray tip. *Analytical Chemistry,* 77(2), pp. 527-533.
Zhang, H. et al., 2013. Detection of low-abundance KRAS mutations in colorectal cancer using microfluidic capillary electrophoresis-based restriction fragment length polymorphism method with optimized assay conditions. *PloS one,* 8(1), p. e54510.
Zhang, Y. et al., 2009. DNA methylation analysis on a droplet-in-oil PCR array. *Lab on a Chip,* 9(8), pp. 1059-1064.
Zhu, Z. et al., 2012. Highly sensitive and quantitative detection of rare pathogens through agarose droplet microfluidic emulsion PCR at the single-cell level. *Lab on a Chip,* 12(20), pp. 3907-3913.